\documentclass[10pt]{article}

\usepackage{epsfig}
\usepackage{cite}
\usepackage{natbib}
\usepackage{amsmath}
\usepackage{amsthm}
\usepackage{easybmat}
\usepackage{float}
\usepackage{lmodern}
\usepackage{lscape,epsfig}
\usepackage{amssymb,amsmath}
\usepackage{multirow}
\usepackage{graphicx}
\usepackage{adjustbox}
\usepackage{bm}
\usepackage{ifxetex,ifluatex}
\usepackage[table,xcdraw]{xcolor}
\usepackage{rotating}
\usepackage{booktabs} 

\usepackage{algorithm}
\usepackage[noend]{algpseudocode}
\usepackage[]{algpseudocode}
\usepackage{algorithmicx}
\usepackage{amsfonts}

\algnewcommand{\IfThenElse}[3]{
  \State \algorithmicif\ #1\ \algorithmicthen\ #2\ \algorithmicelse\ #3}

\algloopdefx{NFor}[1]{\textbf{for all} #1 \textbf{do}}
\algnewcommand\And{\textbf{and}}
\algnewcommand\Or{\textbf{or }}

\DeclareMathOperator*{\argmin}{arg\,min}

\IfFileExists{microtype.sty}{\usepackage{microtype}}{} 
\usepackage[margin=1in]{geometry}
\ifxetex
  \usepackage[setpagesize=false, 
              unicode=false, 
              xetex]{hyperref}
\else
  \usepackage[unicode=true]{hyperref}
\fi
\hypersetup{breaklinks=true,
            bookmarks=true,
            pdfauthor={},
            pdftitle={Additive model for interactions},
            colorlinks=true,
            citecolor=blue,
            urlcolor=blue,
            linkcolor=magenta,
            pdfborder={0 0 0}}
\usepackage{color}
\usepackage{soul} 

\newcommand{\thad}[1]{{\textcolor{green}{#1}}}

\usepackage{color}
\definecolor{eGreen}{rgb}{.057, .549,.065}

\usepackage{titling}
 \title{Functional additive models for optimizing individualized treatment rules}
  \pretitle{\vspace{\droptitle}\centering\huge}
  \posttitle{\par}
  \author{HYUNG G. PARK$^{a\ast}$, EVA PETKOVA$^{a}$, THADDEUS TARPEY$^{a}$, 
R. TODD OGDEN$^{b}$\\[4pt]
\textit{$^{a}$ Division of Biostatistics, Department of Population Health,
New York University} \\ 
\textit{$^{b}$ Department of Biostatistics, Columbia University}
\\[2pt]
} 
  \preauthor{}\postauthor{}
\date{ }

\newtheorem{theorem}{Theorem}

\newtheorem*{notation}{Notation}

\begin{document}
\maketitle
\footnotetext{To whom correspondence should be addressed; parkh15@nyu.edu}


\section*{Abstract} 

A novel functional additive model is proposed which is uniquely modified and constrained to model nonlinear interactions between a treatment indicator and a potentially large number of functional and/or scalar pretreatment covariates. The primary motivation for this approach is to optimize individualized treatment rules based on data from a randomized clinical trial. We generalize functional additive regression models by incorporating treatment-specific components into additive effect components. A structural constraint is imposed on the treatment-specific components in order to provide a class of additive models with main effects and interaction effects that are orthogonal to each other. If primary interest is in the interaction between treatment and the covariates, as is generally the case when optimizing individualized treatment rules, we can thereby circumvent the need to estimate the main effects of the covariates, obviating the need to specify their form and thus avoiding the issue of model misspecification. The methods are illustrated with data from a depression clinical trial with electroencephalogram functional data as patients' pretreatment covariates.

\noindent
{\it Keywords:}  Individualized treatment rules; Functional additive regression; Sparse additive models; Treatment effect-modifiers 

\section{Introduction}\label{sec.introduction} 

We propose a flexible
functional regression approach to optimizing individualized treatment decision 
rules (ITRs) where the treatment has to be chosen to optimize the expected treatment outcome. 
We focus on the situation in which potentially large number of patient characteristics is available as pretreatment functional and/or scalar covariates. 
Recent advances in biomedical imaging, mass spectrometry, and high-throughput gene expression technology produce massive amounts of data on individual patients,  opening up the possibility of tailoring treatments to the biosignatures of individual patients from individual-specific data \citep{McKeague.Qian.2014}. 
Notably, some randomized clinical trials \citep[e.g.,][]{embarc} 
are designed to discover biosignatures  that characterize  patient heterogeneity in treatment responses from vast amounts of patient pretreatment characteristics. 
In this paper, we focus on some specific types of high-dimensional pretreatment patient characteristics 
 observed in the form of curves or images, 
for instance, electroencephalogram (EEG)  measurements. 
 Such data can be viewed as functional \citep[e.g.,][]{fda} and are  becoming increasingly prevalent in modern randomized clinical trials (RCTs) as pretreatment covariates.

Much work has been carried out to develop methods for optimizing ITRs using data from RCTs. 
Regression-based methodologies are intended to
 optimize ITRs by estimating treatment-specific response 
 \citep[e.g.,][]{QianAndMurphy, LU.2011, MC, A.learning.Shi2016, A.learning.Jeng2018,SIMML,Petkova2020}  
while attempting to maintain robustness with respect to model misspecification. 
Machine learning approaches for optimizing ITRs are 
often framed as a classification problem  \citep[e.g.,][]{Zhang.classification, Zhao.2019}, 
including  outcome weighted learning \citep[e.g.,][]{Zhao.2012, Zhao.2015, Song.2015} based on support vector machines, tree-based classification \citep[e.g.,][]{Laber.Zhao.2015}  and adaptive boosting \citep{KANG.2014}, 
among others. 
However, to date there has been relatively little research on ITRs that directly utilize pretreatment functional covariates. 
\cite{McKeague.Qian.2014} proposed methods for optimizing ITRs 
 that depend upon a single pretreatment functional covariate. 
 The flexible functional regression approach of 
 \cite{Ciarleglio.stat} 
  is also restricted to 
  a single pretreatment functional covariate. 
\cite{Ciarleglio.scalar.on.functions}  proposed a method  allowing for 
 multiple functional/scalar covariates, 
and extended to
 incorporate 
a simultaneous covariate selection for ITRs in \cite{Ciarleglio.jrss.c}. However, 
 both of these approaches are limited to a stringent linear model assumption  on the 
  treatment-by-covariates interaction effects that 
  limits flexibility in optimizing ITRs 
  and to two treatment conditions.

In this paper, we allow for
nonlinear interactions between the  treatment and the pretreatment functional covariates on the outcome 
and also for more than  two treatment conditions.  
We 
 incorporate
a simultaneous covariate selection for ITRs through an $L^1$ regularization 
to deal with a large number of functional and/or scalar covariates. 
In a review by \cite{Morris2015} on functional regression, 
two popular approaches to functional additive regression 
are the functional additive 
regression of \cite{FAR} and the functional generalized additive model of \cite{McLean2014}. 
In this paper, we base our method on 
the  functional  additive regression model of \cite{FAR} that utilizes one-dimensional  data-driven functional indices  and the associated additive link functions. 
In \cite{Ciarleglio.stat}, nonlinear effects are presented with 
the functional additive regression of 
  \cite{McLean2014}. 
However, 
the approach of \cite{McLean2014}
 requires more parameters for estimation and is based on an $L^2$ penalty rather than on $L^1$ penalties, 
 which is less suitable  in the context of many functional covariates and when sparsity is desired. 
In this paper, we develop  a 
flexible approach to optimizing ITRs that can easily impose structural constraints  
in modeling nonlinear heterogenous treatment effects with functional and/or scalar pretreatment covariates. 
 


\section{Constrained functional additive models}\label{sec.models} 
We consider 
a treatment response  $Y \in \mathbb{R}$, 
a set of $p$ 
functional-valued pretreatment covariates 
$\bm{X} = (X_1,\ldots, X_p)$, 
and 
$q$
 scalar-valued pretreatment  covariates $\bm{Z} = (Z_1,\ldots, Z_q) \in \mathbb{R}^q$. 
These pretreatment  covariates $(\bm{X}, \bm{Z})$  are considered as potential biomarkers for optimizing ITRs. 
We will assume that each $X_j$ is a 
square integrable random function, defined on a compact interval, say, $[0,1]$, 
without loss of generality.  
Suppose  there are $L$ available treatment options, with treatment  indicator   $A \in \{1,\ldots,L\}$ 
assigned with associated randomization probabilities $(\pi_1,\ldots, \pi_L)$, such that $\sum_{a=1}^{L} \pi_a =1$, $\pi_a > 0$, 
independent of $(\bm{X}, \bm{Z})$.

In this context we focus on optimizing ITRs based on $(\bm{X}, \bm{Z})$.  
For a single decision point, an ITR based on $(\bm{X},\bm{Z})$, 
 which we denote by $\mathcal{D}$, 
 maps a patient with pretreatment characteristics $(\bm{X},\bm{Z})$ to one of the treatment options in $\{1, \ldots, L\}$. 
One popular measure of  the effectiveness of 
 $\mathcal{D}$ is 
the so-called  ``value'' ($V$)  function \citep{Murphy2005}, 
$V(\mathcal{D}) = E[ E[ Y | \bm{X}, \bm{Z}, A= \mathcal{D}(\bm{X},\bm{Z})]]$,  
 the aggregate effect of applying a given treatment regime  $\mathcal{D}$ across the population. 
If we assume, without loss of generality, that a larger value of $Y$ is  better, then the optimal ITR, which we denote as  $\mathcal{D}^{opt}$,  can be defined as $\mathcal{D}$ that maximizes 
$V(\mathcal{D})$. Such a rule 
$\mathcal{D}^{opt}$ 
can be  
shown to satisfy:  
$\mathcal{D}^{opt}(\bm{X},\bm{Z}) =  \operatorname*{arg\,max}_{a \in \{1,\ldots, L\}}   E[ Y | \bm{X},\bm{Z}, A=a ]$. 
In particular, 
$\mathcal{D}^{opt}$ 
does not depend on 
 the  ``main'' effect  of the covariates $(\bm{X},\bm{Z})$ 
 and   depends only on the 
 $(\bm{X},\bm{Z})$-by-$A$ 
   interaction effect \citep{QianAndMurphy} 
   in the mean response function $E[ Y | \bm{X},\bm{Z}, A]$.  However, 
 if this mean response model inadequately represents the interaction effect, the associated ITR may perform poorly.  

Thus, 
we will focus on  modeling 
possibly nonlinear $(\bm{X},\bm{Z})$-by-$A$ interaction effects, 
while allowing for an unspecified main effect of $(\bm{X},\bm{Z})$. 
We base the model on 
 the functional additive 
 model (FAM) of \cite{FAR} allowing for nonlinear $(\bm{X},\bm{Z})$-by-$A$  interactions: 
\begin{equation} \label{the.model}
\begin{aligned}
E[Y | \bm{X}, \bm{Z},A ] =
\underbrace{\mu(\bm{X}, \bm{Z})}_{(\bm{X}, \bm{Z}) \mbox{ ``main'' effect}} + \quad 
\underbrace{ \sum_{j=1}^p g_{j}(\langle X_j, \beta_{j}  \rangle, A ) 
\ + \ \sum_{k=1}^q h_{k}(Z_k, A).}_{(\bm{X}, \bm{Z})\mbox{-by-}A \mbox{ interaction effect}}   
\end{aligned} 
\end{equation}
In model (\ref{the.model}), 
 the treatment 
 $a$-specific (with $a \in \{1,\ldots,L\})$  component functions 
$\{g_{j}(\cdot, a), \ j=1,\ldots, p \}  \cup  \{ h_{k}(\cdot, a),\ k=1,\ldots, q\}$ 
are unspecified 
  smooth 
 one-dimensional (1-D) functions. 
 Specifically, 
each 
function $X_j$ appears 
as a 1-D projection  $\langle X_j, \beta_{j}  \rangle 
 := \int_0^1   X_j(s) \beta_{j}(s) ds$,  
   via the standard $L^2$ inner product with a   
    coefficient function  $\beta_j \in \Theta$, 
where $\Theta$  is the space of square integrable functions   over $[0,1]$, 
 restricted to  a unit $L^2$ norm 
 for model identifiability (due to the unspecified nature of the associated functions $g_{j}(\cdot, a)$). 
  The form of the function $\mu$ in (\ref{the.model}) is left unspecified. 
For model (\ref{the.model}), we assume an additive noise, 
$Y = E[Y | \bm{X}, \bm{Z}, A]  + \epsilon$,  
where  $\epsilon \in \mathbb{R}$  is a zero-mean 
noise with finite variance. 
 
 In model (\ref{the.model}), 
to separate the nonparametric $(\bm{X}, \bm{Z})$ ``main'' effect 
from the additive $(\bm{X}, \bm{Z})$-by-$A$ 
interaction effect components,  
and 
to obtain an identifiable representation, 
we will constrain  
  the $p+q$ component functions $\{g_j, j=1,\ldots,p\} \cup \{h_k, k=1,\ldots,q\}$ associated with the $(\bm{X}, \bm{Z})$-by-$A$  interaction effect 
to satisfy the following identifiability conditions: 
 \begin{equation}   \label{the.condition}
\begin{aligned}
E\big[g_{j}( \langle X_j, \beta_j \rangle, A) \mid X_j \big]  &
 = 0 \quad  
   (\forall \beta_j \in \Theta)
   \quad (j=1,\ldots,p) \quad \mbox{and} \\
E\big[h_{k}( Z_k, A) \mid Z_k \big]   & 
= 0 \quad  
  (k=1,\ldots,q)   
  \end{aligned}
\end{equation}
(almost surely), 
where the expectation is taken with respect to the distribution of $A$ given $X_j$ (or $Z_k$). 
 Condition (\ref{the.condition}) implies  
$E\big[ \sum_{j=1}^p g_{j}( \langle  X_j, \beta_j \rangle, A) 
+ \sum_{k=1}^q h_{k}( Z_k, A)   
 \mid
  \bm{X}, \bm{Z} \big]  =0$ (almost surely), 
which makes not only representation 
(\ref{the.model}) 
identifiable, but also the two effect components in model 
(\ref{the.model}) 
orthogonal to each other. 
We call model (\ref{the.model}) 
subject to the constraint  (\ref{the.condition}), 
a {\em constrained functional additive  model} (CFAM), 
which is the main model of the paper.

 \begin{notation} 
For a fixed $\beta$, 
let us denote 
the   $L^2$ space of component functions, $g(\cdot, \cdot)$, 
over the random variables $(\langle  X, \beta \rangle,A)$ as: 
 $\mathcal{H}^{(\beta)} = \{ g  \mid  E[  g(\langle  X, \beta \rangle,A)] =0,  \lVert g \rVert < \infty \}$,   
with $\lVert g \rVert  =   \sqrt{E\big[  g^{2}(\langle X, \beta \rangle, A)  \big]}$, where the expectation is taken
 with respect to the joint distribution of
  $(\langle  X, \beta \rangle, A)$ 
and  the inner product of the space defined as 
$\langle g, g^\prime \rangle = E[  g(\langle  X, \beta \rangle,A) g^\prime(\langle X, \beta \rangle,A)]$. 
Similarly, let us denote  the  $L^2$ space of  component 
functions, $h(\cdot,\cdot)$,  over  
$(Z,A)$ as: 
$\mathcal{H} = \{ h  \mid  E[  h(Z,A)] =0,  \lVert h  \rVert < \infty \}$ 
with 
  $\lVert h \rVert  =   \sqrt{E\big[  h^{2}(Z,A)  \big]}$, where the expectation is 
   with respect to the 
    distribution of  $(Z,A)$,  
and  similarly defined inner product. 
We suppress the treatment-specific intercepts in models, by removing the treatment $a$-specific  means from $Y$,  
and assume $E[Y|A=a] =0$ $(a=1,\ldots,L)$, i.e., the main effect of  $A$   is $0$, without loss of generality. 
\end{notation}

Under 
the formulation (\ref{the.model}) subject to the constraint (\ref{the.condition}),  
the ``true''  (i.e., optimal) functions, denoted as 
$\{g_j^\ast, j=1,\ldots,p\} \cup \{\beta_j^\ast, j=1,\ldots,p\} \cup \{ h_k^\ast, k=1,\ldots,q\}$ 
 that
  constitute the $(\bm{X},\bm{Z})$-by-$A$ interaction effect,    
 can be viewed as the solution to the constrained optimization: 
 \begin{equation} \label{LS4}
\begin{aligned}
\{ g_{j}^\ast, \beta_{j}^\ast, h_{k}^\ast  \} 
\quad = \quad & \underset{ g_{j}   \in \mathcal{H}_j^{(\beta_j)}, \beta_j \in \Theta,  h_k \in \mathcal{H}_k }{\text{argmin}}
& &E 
 \bigg\{ Y -    \sum_{j=1}^p g_{j}(\langle X_j, \beta_j \rangle, A)  
- \sum_{k=1}^q h_{k}(Z_k,A) 
   \bigg\}^2, 
    \\
& \text{subject to} & & E\left[g_{j}(\langle X_j, \beta_j \rangle,A) | X_j \right]  = 0 \quad 
\forall \beta_j \in \Theta \quad 
(j=1,\ldots,p) \quad \mbox{and}  \\
& 			    & & E\left[h_{k}( Z_k, A) | Z_k \right]  = 0 \quad (k=1,\ldots,q). 
\end{aligned}
\end{equation}
Specifically, 
 representation (\ref{LS4}) 
does not involve the ``main'' effect functional $\mu$, 
due to the orthogonal representation (\ref{the.model}) implied by (\ref{the.condition}). (See Section A.1 of Supporting Information for additional detail.)  
If $\mu$ in  (\ref{the.model})  is a complicated functional subject to model misspecification, 
exploiting the representation 
on the right-hand side of 
(\ref{LS4})   
for   $\{g_j^\ast, j=1,\ldots,p\} \cup \{\beta_j^\ast, j=1,\ldots,p\} \cup \{ h_k^\ast, k=1,\ldots,q\}$ 
 on the left-hand side  
 is  particularly appealing, 
  as it provides a means of estimating 
 the interaction terms 
   without having to specify $\mu$, thereby avoiding 
   any issue of 
   possible model misspecification for $\mu$. 
The function $\mu$ can also be 
specified  similar to 
 (\ref{LS4}) 
and estimated separately (see Section A.6 of Supporting Information), due to orthogonality in model (\ref{the.model}). 
In particular, estimators of $\{g_j^\ast , \beta_j^\ast, h_k^\ast \}$  based on  optimization (\ref{LS4}) can be improved in terms of efficiency
if  $Y$  in (\ref{LS4}) is replaced by a ``residualized'' response 
$Y - \widehat{\mu}(\bm{X}, \bm{Z})$, where $\widehat{\mu}$ is some estimate of $\mu$  (see also Section A.6 of Supporting Information). 
However, for simplicity, we will focus on the representation  (\ref{LS4}) with the ``unresidualized''  $Y$.

  Under model (\ref{the.model}), 
  the potential treatment effect-modifying variables among 
   $\{ X_j, j=1,\ldots,p  \} \cup \{ Z_k, k=1,\ldots,q \}$  appear in 
  the model, 
  only through
 the interaction effect terms in  (\ref{the.model}) that specify 
and characterize  the heterogeneous treatment effects. 
  \cite{SAM} proposed a sparse additive model (SAM) 
   for relevant covariate selection in a high-dimensional additive regression. 
 As in SAM, to deal with a large $p+q$ and to achieve treatment effect-modifying variable selection, we  impose sparsity  
 on the set of component functions $\{ g_{j}, j=1,\ldots,p \} \cup \{ h_{k}, k=1,\ldots,q \}$ of CFAM (\ref{the.model}), 
 under the often reasonable assumption that most covariates are inconsequential as treatment effect-modifiers.   
This sparsity structure on 
the set of component functions 
can be usefully incorporated into the optimization-based representation (\ref{LS4}) for $\{ g_{j}^\ast, \beta_{j}^\ast, h_{k}^\ast  \}$:   
 \begin{equation} \label{LS5}
\begin{aligned}
\{ g_{j}^\ast, \beta_{j}^\ast, h_{k}^\ast  \} 
= 
& \underset{ g_{j}   \in \mathcal{H}_j^{(\beta_j)}, \beta_j \in \Theta,  h_k \in \mathcal{H}_k }{\text{argmin}}
& &E 
\bigg\{ Y -    \sum_{j=1}^p g_{j}(\langle X_j, \beta_j \rangle,A)  
- \sum_{k=1}^q h_{k}(Z_k,A) 
   \bigg\}^2 
   + 
     \lambda \bigg\{ \sum_{j=1}^p \lVert g_{j} \rVert + \sum_{k=1}^q \lVert h_{k} \rVert \bigg\},
    \\
& \text{subject to} & & E[g_{j}(\langle X_j, \beta_j \rangle,A) | X_j ]  = 0 \quad 
\forall \beta_j \in \Theta \quad 
(j=1,\ldots,p) \quad \mbox{and}  \\
& 			    & & E[h_{k}(Z_k,A) | Z_k ]  = 0 \quad (k=1,\ldots,q),  
\end{aligned}
\end{equation}
for some sparsity-inducing parameter $\lambda  \ge 0$. 
The term $\sum_{j=1}^p \lVert g_{j} \rVert + \sum_{k=1}^q \lVert h_{k} \rVert $ in (\ref{LS5})  behaves like an $L^1$ \textit{ball} across different functional components $\{ g_j, j=1,\ldots,p; \ h_k, k=1,\ldots,q \}$ to encourage   functional 
sparsity. 
For example, a relatively large value of  $\lambda$ in 
(\ref{LS5})  will result in many components to be exactly zero, thereby enforcing 
 sparsity on the set of functions 
 $\{ g_{j}^\ast, h_k^\ast\}$ 
  on the left-hand side of (\ref{LS5}). Specifically, 
equation (\ref{LS5}) can help model selection when dealing with potentially many functional/scalar pretreatment covariates.

  \section{Estimation} \label{sec.estimation}

We first consider a population characterization of the algorithm  for solving (\ref{LS5}) in Section~\ref{sec.population.algorithm} 
and then a sample counterpart of the population algorithm in Section~\ref{sec.sample.algorithm}.

\subsection{Population algorithm}  \label{sec.population.algorithm}
 
For a set of fixed  coefficient functions $\{ \beta_j, j=1,\ldots,p \}$,  
 the minimizing component function $g_j \in \mathcal{H}_j^{(\beta_j)}$ 
  (and $h_k \in \mathcal{H}_k$) 
 for each $j$ (and each $k$)  of the  constrained objective function of (\ref{LS5}) 
has  a component-wise closed-form expression.

\begin{theorem} \label{theorem1}
Given $\lambda \ge 0$ 
and a set of fixed single-index coefficient functions 
$\{ \beta_j, j=1,\ldots,p \}$,  
 the minimizing component function  $g_j \in \mathcal{H}_j^{(\beta_j)}$ 
of the constrained objective function of (\ref{LS5})  satisfies: 
\begin{equation}\label{g.solution}
g_{j}(\langle X_j, \beta_j \rangle,A) \ =\  \left[ 1- \frac{\lambda}{\lVert f_{j} \rVert} \right]_{+} f_{j}(\langle X_j, \beta_j \rangle,A)  \quad\ \mbox{(almost surely)},  
\end{equation} 
where the function $f_j \in \mathcal{H}_j^{(\beta_j)}$: 
\begin{equation}\label{proj.1}
f_{j}(\langle X_j, \beta_j \rangle,A)  \ :=  \ E[R_{j} |\langle X_j, \beta_j \rangle,A ] \ -  \ E[R_{j} | \langle X_j, \beta_j \rangle ],  
\end{equation}
in which 
\begin{equation}\label{partial.residual}
R_j = Y - \sum_{ j^\prime \ne j}g_{j^\prime}(\langle X_{j^\prime}, \beta_{j^\prime} \rangle,A) - \sum_{k=1}^q h_{k,A}(Z_k)
 \end{equation}
represents  the $j$th (functional covariate's) partial residual; 
similarly, 
the minimizing component function $h_k \in \mathcal{H}_k$ of the constrained objective function of (\ref{LS5})   satisfies: 
\begin{equation}\label{h.solution}
h_{k}(Z_k,A) \ =\  \left[ 1- \frac{\lambda}{\lVert \check{f}_{k} \rVert} \right]_{+}  \check{f}_{k}(Z_k,A)  \quad \  \mbox{(almost surely)},
\end{equation} 
where the function $\check{f}_k \in \mathcal{H}_k$:
\begin{equation}\label{proj.2}
\check{f}_{k}(Z_k,A)  \ :=  \ E[\check{R}_{k} |Z_k,A] \ -  \ E[\check{R}_{k} | Z_k],  
\end{equation}
and 
\begin{equation}\label{partial.residual2}
\check{R}_k  = Y - \sum_{j=1}^p g_{j}(\langle X_{j}, \beta_{j} \rangle,A) - \sum_{k^\prime \ne k} h_{k^\prime}(Z_{k^\prime},A)
 \end{equation}
represents the  $k$th (scalar covariate's) partial residual.  (In (\ref{g.solution}) and  (\ref{h.solution}), 
 $[u]_{+} = \max(0,u)$ represents the positive part of $u$.)
\end{theorem}

The proof of Theorem~\ref{theorem1} is in Section A.2 of Supporting Information. 
Given a sparsity tuning parameter $\lambda \ge 0$, 
optimization (\ref{LS5}) can be split into two iterative steps \citep{FRAME, FAR}. 
First (\textit{Step 1}), for a set of   fixed single-indices $\langle X_{j}, \beta_{j} \rangle$  $(j=1,\ldots,p)$, 
the component functions 
$\{g_j, j=1,\ldots,p \} \cup \{h_k, k=1,\ldots, q\}$ of the model 
can be found by a coordinate descent procedure 
that fixes 
$\{g_{j'}; j' \ne  j \} \cup \{ h_k, k=1,\ldots,q\}$ 
and obtains $g_{j}$ 
by equation (\ref{g.solution}) 
(and that fixes $\{ g_j, j=1,\ldots,p\} \cup \{h_{k'}; k' \ne  k \}$ and obtains  $h_k$ by equation   (\ref{h.solution})), 
and then iterates  through all $j$ and $k$ until convergence. This step (\textit{Step 1}) amounts to fitting a  SAM  \citep{SAM}  subject to the constraint  (\ref{the.condition}). 
Second (\textit{Step 2}), for a set of fixed component functions 
$\{g_{j}, j=1,\ldots,p \} \cup \{ h_k, k=1,\ldots,q\}$, the $j$th single-index coefficient function $\beta_j \in \Theta$ 
can be optimized by solving, 
for each $j \in \{1,\ldots, p\}$ separately: 
\begin{equation}\label{the.MSE.CFAM}
\underset{ \beta_j \in \Theta}{\text{minimize}} \quad   E 
 \bigg\{ R_j -  g_{j}( \langle X_{j}, \beta_{j} \rangle, A)   \bigg\}^2 
  \quad (j=1,\ldots,p),  
\end{equation}
where the $j$th partial residual $R_j$ is defined  in (\ref{partial.residual}). 
These two steps can be iterated until convergence  to obtain a population solution $\{ g_{j}^\ast, \beta_{j}^\ast, h_{k}^\ast  \}$ on the left-hand side of (\ref{LS5}).

To obtain a sample version of the population solution, we can insert sample estimates into the population algorithm,  
 as in standard backfitting in estimating generalized additive models \citep{GAM}, which we describe in the next subsection.

\subsection{Sample version of the population algorithm} \label{sec.sample.algorithm}

To simplify the exposition, 
we only describe 
 the optimization of 
  $g_{j}(\langle X_{j}, \beta_{j} \rangle,A)$  $(j=1,\ldots,p)$ 
associated with the functional covariates $X_j$ $(j=1,\ldots, p)$. 
The components $h_{k}(Z_k,A)$  $(k=1,\ldots,q)$ 
associated with the scalar covariates $Z_k$ $(k=1,\ldots, q)$ in (\ref{LS5})  
are optimized in the same way, except that 
we do not need to perform  \textit{Step 2} of the alternating optimization procedure; 
i.e.,
when optimizing  $h_{k}(Z_k,A)$  $(k=1,\ldots,q)$, 
we only perform \textit{Step 1}.

  \subsubsection{Step 1} \label{step1}

First, we consider  a sample version of \textit{Step 1} of the population algorithm.  
Suppose we are given a set of estimates $\{ \widehat{\beta}_{j}, j=1,\ldots, p\}$ 
and the data-version of the $j$th partial residual $R_j$ in (\ref{partial.residual}): 
$\widehat{R}_{ij} = Y_{i} - \sum_{ j^\prime \ne j}\widehat{g}_{j^\prime}(  \langle X_{ij'}, \widehat{\beta}_{j'} \rangle, A_i ) - \sum_{k=1}^q \widehat{h}_k(Z_{ik},A_i)$ 
$(i=1,\ldots,n)$,  
where $\widehat{g}_{j^\prime}$ represents a current estimate for $g_{j^\prime}$ 
and $\widehat{h}_{k}$ that for $h_{k}$. 
For each $j$, 
we update the component function $g_j$ in (\ref{g.solution}) in two steps: 
first, estimate  the function $f_j$ in 
 (\ref{proj.1});  
second, plug the estimate of $f_j$ 
into $\left[ 1- \frac{\lambda}{\lVert f_{j} \rVert} \right]_{+}$    
 in (\ref{g.solution}),  
to obtain the soft-thresholded estimate $\widehat{g}_j$.

Although any linear smoothers can be utilized to obtain estimators $\{ \widehat{g}_j, j=1,\ldots,p \}$ (see Section A.3 of Supporting Information), 
we shall focus on regression spline-type estimators, 
which are 
simple and computationally efficient to implement. For each $j$ and 
$\beta_j = \widehat{\beta}_j,$ 
we will 
 represent 
 the component function 
 $g_j \in \mathcal{H}_j^{(\widehat{\beta}_j)}$ 
on the right-hand side of  (\ref{LS5}) as: 
\begin{equation} \label{eq.5}
\begin{aligned}
g_{j}(\langle X_{j}, \widehat{\beta}_{j} \rangle,a)  = \bm{\Psi}_j(\langle X_{j}, \widehat{\beta}_{j} \rangle)^\top \bm{\theta}_{j,a}   \quad (a=1,\ldots,L) 
\end{aligned}
\end{equation}
for some prespecified $d_j$-dimensional basis $\bm{\Psi}_j(\cdot)$ 
(e.g., cubic $B$-spline basis with $d_j - 4$ interior knots, 
evenly placed over the range  (scaled to, say, $[0,1]$) of the observed values of $\langle X_{j}, \widehat{\beta}_{j} \rangle$)
and a set of unknown  treatment $a$-specific basis coefficients $\{ \bm{\theta}_{j,a}  \in \mathbb{R}^{d_j} 
\}_{a \in \{1,\ldots,L\}}$. 
Based on representation (\ref{eq.5}) of $g_j \in \mathcal{H}_j^{(\widehat{\beta}_j)}$ for fixed $\widehat{\beta}_j$,  
the constraint  
$E[g_{j}(\langle X_{j}, \beta_{j} \rangle,A) | X_j] 
= 0$ 
in (\ref{LS5}) on $g_j$, 
for fixed $\beta_j = \widehat{\beta}_j,$ 
can be simplified to:  
$E[ \bm{\theta}_{j,A}  ] = \sum_{a=1}^{L} \pi_a \bm{\theta}_{j,a} = \bm{0}$. 
If we fix $\beta_j = \widehat{\beta}_j$, 
the constraint in (\ref{LS5}) on the function $g_j$
 can then be succinctly written in matrix form: 
\begin{equation}\label{lin.constr}
\bm{\pi}^{(j)} \bm{\theta}_j  = \bm{0},  
\end{equation} 
where  $\bm{\theta}_j := (\bm{\theta}_{j,1}^\top, \bm{\theta}_{j,2}^\top, \ldots, \bm{\theta}_{j,L}^\top )^\top \in \mathbb{R}^{d_jL}$  
is the vectorized version of the basis coefficients $\{\bm{\theta}_{j,a}\}_{a \in \{1,\ldots,L\}}$, 
 and the $d_j \times d_jL$  
 matrix  $\bm{\pi}^{(j)} := (\pi_1 \bm{I}_{d_j}; \pi_2 \bm{I}_{d_j};  \ldots; \pi_L \bm{I}_{d_j})$
where $\bm{I}_{d_j}$ is the $d_j \times d_j$ identity matrix.

Let the $n \times d_j$ matrices 
 $\bm{D}_{j,a}$ $(a=1,\ldots,L)$ 
  denote the evaluation matrices of the basis $\bm{\Psi}_j(\cdot)$ 
  on 
 $\langle X_{ij}, \widehat{\beta}_{j} \rangle$ 
 $(i=1,\ldots,n)$ specific to the treatment $A=a$ $(a=1,\ldots,L)$,
 whose $i$th row is  the $1 \times d_j$ vector $\bm{\Psi}_j(\langle X_{ij}, \widehat{\beta}_{j} \rangle)^\top$  if  $A_i = a$, and a row of zeros $\bm{0}^\top$ if $A_i \ne a$. 
Then the column-wise concatenation 
of the design matrices $\{ \bm{D}_{j,a}\}_{a \in \{1,\ldots,L\}}$, i.e., the $n \times d_jL $ matrix
 $\bm{D}_j  = (\bm{D}_{j,1}; \bm{D}_{j,2}; \ldots; \bm{D}_{j,L})$,  
 defines the model matrix associated with the vectorized basis coefficient 
  $\bm{\theta}_j  \in \mathbb{R}^{d_jL}$, 
  vectorized across $\{ \bm{\theta}_{j,a}  
\}_{a \in \{1,\ldots,L\}}$ in representation (\ref{eq.5}).  
 We can then  represent 
  $g_{j}(\langle X_{j}, \widehat{\beta}_{j} \rangle,A)$  
of (\ref{eq.5}), based on the sample data, by the length-$n$ vector: 
\begin{equation} \label{gj.vector1}
\bm{g}_j  = \bm{D}_j \bm{\theta}_j \in \mathbb{R}^n 
\end{equation}
subject to the linear constraint (\ref{lin.constr}) on the parameters $\bm{\theta}_j$. 
(Similarly, we can represent $h_{k}(Z_k,A)$ by a length-$n$ vector.)

The linear constraint in (\ref{lin.constr}) 
on $\bm{\theta}_j$ can be conveniently absorbed into the model matrix $\bm{D}_j$ in (\ref{gj.vector1}) 
by reparametrization, which we describe next. 
We can find a $d_jL \times d_j(L-1)$ basis matrix $\bm{n}^{(j)}$  (that spans  the \textit{null} space of the linear constraint (\ref{lin.constr})), 
such that,  if we set $\bm{\theta}_j = \bm{n}^{(j)} \widetilde{\bm{\theta}}_j$ 
for any arbitrary vector $\widetilde{\bm{\theta}}_j \in \mathbb{R}^{d_j(L-1)}$, then the vector  $\bm{\theta}_j \in \mathbb{R}^{d_j L}$ automatically satisfies the constraint (\ref{lin.constr}): 
$\bm{\pi}^{(j)} \bm{\theta}_j  = \bm{0}$. 
Such a basis matrix $\bm{n}^{(j)}$ 
can be constructed by a QR decomposition of the matrix 
$\bm{\pi}^{(j)\top}$. 
Then representation (\ref{gj.vector1}) can be reparametrized, 
in terms of the unconstrained $\widetilde{\bm{\theta}}_j \in \mathbb{R}^{d_j(L-1)}$ 
by replacing $\bm{D}_j$ 
 in (\ref{gj.vector1}) with a reparametrized model matrix 
$\widetilde{\bm{D}}_j =  \bm{D}_j  \bm{n}^{(j)}$: 
\begin{equation} \label{gj.vector2}
\begin{aligned}
\bm{g}_j = \widetilde{\bm{D}}_j \widetilde{\bm{\theta}}_j .
\end{aligned}
\end{equation} 
Theorem~\ref{theorem1}, together with  Section A.4 of Supporting Information, 
 indicates that (for fixed $\beta_j = \widehat{\beta}_j$) 
the coordinate-wise minimizing function $g_j$ 
of the right-hand side of (\ref{LS5}) 
can be estimated based on the sample by:   
\begin{equation} \label{ghat.solution}
\begin{aligned}
\widehat{\bm{g}}_j  = \left[ 1- \frac{\lambda}{\sqrt{ \frac{1}{n} \lVert  \widehat{\bm{f}}_j \rVert^2} } \right]_{+}  \widehat{\bm{f}}_j  \quad 
\end{aligned}
\end{equation}
where 
\begin{equation} \label{f.solution}
\widehat{\bm{f}}_j  = \widetilde{\bm{D}}_j (\widetilde{\bm{D}}_j^{\top} \widetilde{\bm{D}}_j)^{-1} \widetilde{\bm{D}}_j^{\top} \widehat{\bm{R}}_j,    
\end{equation}
in which $\widehat{\bm{R}}_j  = \bm{Y} -  \sum_{j^\prime \ne j} \widehat{\bm{g}}_{j^\prime} - \sum_{k=1}^q \widehat{\bm{h}}_{k}$   
corresponds  to the estimated $j$th 
partial residual vector. 
(Similarly, we can represent the coordinate-wise minimizing function $h_k$  in (\ref{h.solution}), based on the observed data by 
a length-$n$ vector $\widehat{\bm{h}}_k$.) 
If we set each $\beta_j$ equal to its corresponding estimate $\widehat{\beta}_j$  $(j=1,\ldots,p)$, then  
based on the sample counterpart (\ref{ghat.solution}) of the coordinate-wise solution (\ref{g.solution}), a highly efficient coordinate descent algorithm can be conducted to  optimize $\{g_j, j=1,\dots,p\} \cup \{h_k,k=1,\ldots,q\}$  simultaneously. 
Let $\widehat{s}_j^{(\lambda)} := \left[ 1-  \lambda  \sqrt{n}/  \lVert  \widehat{\bm{f}}_j \rVert  \right]_{+}$ in (\ref{ghat.solution}) denote the soft-threshold shrinkage factor associated with the un-shrunk estimate $\widehat{\bm{f}}_j$ in (\ref{f.solution}). 
At convergence of the coordinate descent, 
we obtain a basis coefficient estimate of $\widetilde{\bm{\theta}}_j$ associated with representation (\ref{gj.vector2}): 
\begin{equation} \label{coef.hat}
 \widehat{\widetilde{\bm{\theta}}}_j =  \widehat{s}_j^{(\lambda)}  (\widetilde{\bm{D}}_j^{\top} \widetilde{\bm{D}}_j)^{-1} \widetilde{\bm{D}}_j^{\top} \widehat{\bm{R}}_j,   
\end{equation}
which in turn implies an estimate of $\bm{\theta}_j$ in (\ref{gj.vector1}):  
$\widehat{\bm{\theta}}_j = (\widehat{\bm{\theta}}_{j,1}^\top,  \widehat{\bm{\theta}}_{j,2}^\top, \ldots, \widehat{\bm{\theta}}_{j,L}^\top   )^\top = \bm{n}^{(j)} \widehat{\widetilde{\bm{\theta}}}_j$. 
Specifically, 
this gives an estimate of the treatment $a$-specific function $g_{j}(\cdot,a)$ 
$(a=1,\ldots,L)$ in model (\ref{the.model}): 
\begin{equation} \label{g.hat.estimate}
\begin{aligned}
\widehat{g}_{j}(\cdot,a) = \bm{\Psi}_j(\cdot)^\top  \widehat{\bm{\theta}}_{j,a} \quad (a=1,\ldots,L) 
\end{aligned} 
\end{equation}
 estimated within the class of functions (\ref{eq.5}),  
for a  given  tuning parameter  $\lambda \ge 0$ controlling 
  the soft-threshold shrinkage factor $\widehat{s}_j^{(\lambda)}$     in (\ref{coef.hat}),   
 resulting in the functions 
   $\{ \widehat{g}_j, j=1,\ldots, p \} \cup \{ \widehat{h}_k, k=1,\ldots, q\}$; this completes Step 1 of the alternating optimization procedure.

  \subsubsection{Step 2}\label{step2}

We now consider a sample version of \textit{Step 2} of the population algorithm 
that optimizes 
 the coefficient functions $\{ \beta_j, j=1,\ldots, p\}$ on the right-hand side of (\ref{LS5}), 
  for a fixed set of the 
component function estimates 
  $\{\widehat{g}_{j}, j=1,\ldots,p \} \cup \{ \widehat{h}_k, k=1,\ldots,q\}$ 
provided by \textit{Step 1}. 
 As an empirical approximation to 
   (\ref{the.MSE.CFAM}), we consider 
 \begin{equation}\label{sample.mse}
 \underset{ \beta_{j} \in \Theta }{\text{minimize}} \quad 
 \sum_{i=1}^n  \left\{ \widehat{R}_{ij} -  \widehat{g}_{j} (\langle X_{ij}, \beta_{j} \rangle,A_i)  \right\}^2 \quad (j=1,\ldots,p), 
\end{equation} 
 where $\widehat{R}_{ij}$ is the $i$th element of $\widehat{\bm{R}}_j \in \mathbb{R}^n$ in (\ref{ghat.solution}).    
For this iterative estimation step, 
 solving  (\ref{sample.mse}) 
can be approximately achieved based on a first-order Taylor series 
 approximation of the term 
$\widehat{g}_{j}(\langle X_{ij}, \beta_{j} \rangle, A_i)$ 
at the current estimate, which we denote as $\widehat{\beta}_{j}^{(c)} \in \Theta$: 
\begin{equation} \label{Q.hat2}
\begin{aligned}
 \sum_{i=1}^n  \left\{ \widehat{R}_{ij} -  \widehat{g}_{j}\left(\langle X_{ij}, \beta_{j} \rangle,A_i\right)  \right\}^2  \ 
& \approx  \    \sum_{i=1}^n \left\{ \widehat{R}_{ij}  -  \widehat{g}_{j}(\langle X_{ij}, \widehat{\beta}_{j}^{(c)} \rangle,A_i)   
-   \dot{\widehat{g}}_{j}(\langle X_{ij}, \widehat{\beta}_{j}^{(c)} \rangle,A_i) \ \langle X_{ij}, \beta_j -  \widehat{\beta}_{j}^{(c)} \rangle 
\right\}^2  \\ 
&= \    \sum_{i=1}^n   \left\{ \widehat{R}_{ij}^{\ast} -    
 \langle X_{ij}^\ast, \beta_j \rangle 
  \right\}^2, 
\end{aligned}
\end{equation}
where the 
``modified'' residuals $\widehat{R}_{ij}^{\ast}$ and  the ``modified'' 
covariates $X_{ij}^\ast$ are defined as: 
\begin{equation} \label{modified.variables}
\begin{aligned}
\widehat{R}_{ij}^{\ast}  &= \widehat{R}_{ij} -  \widehat{g}_{j}(\langle X_{ij}, \widehat{\beta}_{j}^{(c)} \rangle, A_i)  +   \dot{\widehat{g}}_{j}(\langle X_{ij}, \widehat{\beta}_{j}^{(c)} \rangle, A_i) \ \langle X_{ij}, \widehat{\beta}_{j}^{(c)} \rangle  \quad &  (i=1,\ldots,n ), \\
X_{ij}^{\ast} &=  \dot{\widehat{g}}_{j}(\langle X_{ij}, \widehat{\beta}_{j}^{(c)} \rangle, A_i)  \ X_{ij}  &  (i=1,\ldots,n ), 
\end{aligned}
\end{equation}
in which each $\dot{\widehat{g}}_{j}(\cdot,a)$ denotes 
the first derivative  of $\widehat{g}_{j}(\cdot, a)$ in (\ref{g.hat.estimate}) given by \textit{Step 1}.  
We can perform a functional linear regression 
\citep[e.g.,][]{spline.flm} with scalar response $\widehat{R}_{ij}^{\ast}$ 
 and (functional) covariate $X_{ij}^{\ast}$
 to minimize  the right-hand side of  (\ref{Q.hat2})  over $\beta_j \in \Theta$.  
Specifically, 
 the smooth coefficient function $\beta_j$ in  (\ref{Q.hat2}) 
is  represented by  
a prespecified and normalized 
$m_j$-dimensional $B$-spline basis  $B_j(s) = (b_{j1}(s), \ldots, b_{jm_j}(s))^\top \in \mathbb{R}^{m_j}$,  
where 
 $m_j$ depends  only on the sample size $n$ \citep{FAR}:   
\begin{equation} \label{p.spline} 
\beta_j(s) = \sum_{r=1}^{m_j} b_{jr}(s) \gamma_{jr} \quad s \in [0,1], 
\end{equation} 
with an unknown basis coefficient vector 
$\bm{\gamma}_j = (\gamma_{j1}, \gamma_{j2}, \ldots, \gamma_{jm_j})^\top \in \mathbb{R}^{m_j}$. 
Suppose the function $X_{ij}$ $(i=1,\ldots,n)$ is 
 discretized at 
points $\{ s_l : 0 = s_1 < s_2< \ldots < s_{r_j}  = 1 \}$. 
Using the approximation 
$\langle X_{ij}, \widehat{\beta}_{j}^{(c)} \rangle \approx  
 \sum_{l=1}^{r_j} \Delta_l X_{ij}(s_l) \widehat{\beta}_{j}^{(c)}(s_l)$ 
where $\Delta_l$ is the distance between two neighboring discretization points, 
we 
approximate 
$\widehat{R}_{ij}^{\ast}$ 
and $X_{ij}^\ast$ in (\ref{modified.variables}). 
Let  $\bm{X}_j^\ast$ be the $n \times r_j$ matrix whose $i$th row is the discretized function $X_{ij}^{\ast}(s_{l})$ $(l=1,\ldots,r_j)$, and 
$\bm{B}_j$  the $r_j \times m_j$ matrix whose $l$th row 
is the  evaluated basis $B_j(s) \in \mathbb{R}^{m_j}$  at the $l$th point $s = s_l$ $(l=1,\ldots,r_j)$.
Given  $\beta_j(s)$ 
  discretized at the points $s = s_l$ $(l=1,\ldots, r_j)$  in (\ref{p.spline}), 
 we can represent the right-hand side of  (\ref{Q.hat2}) as: 
 \begin{equation} \label{flm}
    \lVert  \bm{R}_j^\ast   - \bm{U}_j^\ast \bm{\gamma}_j \rVert^2, 
  \end{equation} 
  where $\bm{R}_j^\ast  := (R_{1j}^\ast, \ldots, R_{nj}^\ast)^\top \in \mathbb{R}^{n}$ 
  and $\bm{U}_j^\ast   := \Delta \bm{X}_j^\ast \bm{B}_j$. 
  Minimizing 
(\ref{flm}) 
over $\bm{\gamma}_j \in \mathbb{R}^{d}$   for each $j$ separately ($j=1,\ldots, p$) 
provides 
estimates 
$\{ \widehat{\beta}_j, j=1,\ldots, p \}$ of the coefficient  functions; 
here, the 
 minimizer $\widehat{\bm{\gamma}}_j$ 
for  (\ref{flm})  
 is scaled to 
  $\lVert \widehat{\bm{\gamma}}_j \rVert =1$, 
so that the resulting
  $\widehat{\beta}_j(s) = \sum_{r=1}^{m_j} b_{jr}(s) \widehat{\gamma}_{jr}$ $(s \in [0,1])$ 
  satisfies the identifiability constraint 
 $\widehat{\beta}_j \in \Theta$. 
 This completes \textit{Step 2} of the alternating optimization procedure.


\subsubsection{Initialization and convergence criterion}

At the initial iteration,    we need some estimates $\{ \widehat{\beta}_j, j=1,\ldots,p \}$ of the single-index coefficient functions 
to initialize the single-indices $\{ u_j = \langle \widehat{\beta}_j, X_j \rangle, j=1,\ldots,p \}$, 
in order to perform 
\textit{Step 1} (i.e., the coordinate-descent procedure) of the estimation procedure described in Section~\ref{step1}. 
At the initial iteration, we take $\widehat{\beta}_j(s) =1$ ($s \in [0,1]$), i.e., we take $u_j = \int_0^1 X_j(s) ds$ $(j=1,\ldots,p)$, which corresponds to 
the common practice of taking a na\"ive scalar summary of each functional covariate. 
 The proposed algorithm alternating between \textit{Step 1} and  \textit{Step 2} 
  terminates when the estimates $\{ \widehat{\beta}_j, j=1,\ldots,p \}$ converge. 
  To be specific, 
  the algorithm terminates when $\max_{j=1,\ldots,p,r=1,\ldots,m_j}  \lVert (\widehat{\gamma}_{jr}  - \widehat{\gamma}_{jr}^{(c)}) / \widehat{\gamma}_{jr} \rVert$ 
 is less than a prespecified convergence tolerance; here, $\widehat{\gamma}_{jr}^{(c)}$ represents the current estimate for  $\gamma_{jr}$ in (\ref{p.spline}) at the beginning of \textit{Step 1}, and $\widehat{\gamma}_{jr}$ is the estimate at the end of \textit{Step 2}. 
  We summarize the computational procedure  in Algorithm~\ref{algorithm1}.

\begin{algorithm} 
\caption{Estimation of constrained functional additive models} \label{algorithm1}
\begin{algorithmic} [1]
 \State  \textbf{Input}: Data $\bm{Y} \in \mathbb{R}^n$, $\bm{A} \in \mathbb{R}^n$, $\bm{X}_j \in \mathbb{R}^n \times \mathbb{R}^{r_j}$ $(j=1,\ldots,p)$, 
 and 
  $\lambda \ge 0$ 
   \State  \textbf{Output}:  Estimated functions 
   $\{ \widehat{\beta}_j, j=1,\ldots, p\}$ and $\{ \widehat{g}_j, j=1,\ldots,p  \}$ 
   \State Initialize  $\widehat{\beta}_j(s)=1$ $(s \in [0,1])$  $(j=1,\ldots,p)$.  
\While  
{until convergence of $\{ \widehat{\beta}_j, j=1,\ldots,p \}$,}
 {iteratate between Step 1 and Step 2:}  \\
   $\langle$Step 1$\rangle$

   \State  Fix  $\{ \widehat{\beta}_j, j=1,\ldots,p \}$, and compute 
   $ \widetilde{\bm{D}}_j (\widetilde{\bm{D}}_j^{\top} \widetilde{\bm{D}}_j)^{-1} \widetilde{\bm{D}}_j^{\top} $ in (\ref{f.solution})   $(j=1,\ldots,p)$.   
   \State Initialize $\widehat{\bm{g}}_j = \bm{0} \in \mathbb{R}^n$ $(j=1,\ldots,p)$.  
  \While 
    {until convergence of 
    $\{ \widehat{\bm{g}}_j, j=1,\ldots,p\}$,
    } {iterate through $j = 1,\ldots, p:$} 
   \State Compute the partial residual $\widehat{\bm{R}}_j  = \bm{Y} -  \sum_{j^\prime \ne j} \widehat{\bm{g}}_{j^\prime}$. 
        \State  Compute $\widehat{\bm{f}}_j$ in (\ref{f.solution}); then compute the thresholded estimate $\widehat{\bm{g}}_j$ in  (\ref{ghat.solution}). 
 \EndWhile
     \\ 
     $\langle$Step 2$\rangle$
        \State  Fix $\{ \widehat{g}_j, j=1,\ldots,p\}$ in (\ref{g.hat.estimate}), 
        and solve (\ref{sample.mse}) based on (\ref{flm}); update $\widehat{\beta}_j$
        $(j=1,\ldots,p)$. 
\EndWhile
\end{algorithmic}
\end{algorithm}

   In Algorithm~\ref{algorithm1}, 
  if the $j$th soft-threshold  shrinkage factor 
  $\widehat{s}_j^{(\lambda)}= \left[ 1-  \lambda  \sqrt{n}/  \lVert  \widehat{\bm{f}}_j \rVert  \right]_{+} $ 
  in (\ref{ghat.solution}) 
  is $0$, then the associated $X_j$ is absent from the model. 
  Therefore, the corresponding coefficient function $\widehat{\beta}_j$ will not be updated, 
  and this greatly reduces the computational cost  when 
  most of the shrinkage factors $\widehat{s}_j^{(\lambda)}$ are zeros. 
In  Algorithm~\ref{algorithm1},  
the smoother matrix 
$ \widetilde{\bm{D}}_j (\widetilde{\bm{D}}_j^{\top} \widetilde{\bm{D}}_j)^{-1} \widetilde{\bm{D}}_j^{\top} $ in (\ref{ghat.solution})   $(j=1,\ldots,p)$ needs to be computed only once 
at the beginning of \textit{Step 1} given fixed $\{ \widehat{\beta}_j, j=1,\ldots,p \}$, 
and therefore the coordinate-descent updates in \textit{Step 1} can be performed very efficiently \citep{FRAME}.   
The sparsity tuning parameter $\lambda \ge 0$ can be chosen to minimize
an estimate of the expected squared error of the estimated models 
over a dense grid of $\lambda$'s,  estimated, for example, by a 10-fold cross-validation.

\section{Simulation study}\label{sec.simulation} 

\subsection{ITR  estimation performance} \label{ITR.sim}
In this section, we assess the optimal ITR estimation performance of the proposed method based on  simulations. 
We generate  $n$ independent copies of  $p$ functional-valued covariates $\bm{X}_i = (X_{i1}, X_{i2}, \ldots, X_{ip})$  $(i=1,\ldots, n)$, 
where 
we use a $4$-dimensional Fourier basis,  $\bm{\Phi}(s) =  (\sqrt{2} \sin(2 \pi s),  \sqrt{2} \cos(2 \pi s),  \sqrt{2} \sin(4 \pi s), \sqrt{2} \cos(4 \pi s) )^\top \in \mathbb{R}^4$ $(s \in [0,1])$, 
and random coefficients 
$\widetilde{\bm{x}}_{ij}  \in \mathbb{R}^4$, 
each independently  following 
$\mathcal{N}(\bm{0}, \bm{I}_4)$,
to form the functions 
$X_{ij}(s) = \bm{\Phi}(s)^\top \widetilde{\bm{x}}_{ij}$ 
$(s \in [0,1])$  $(i=1,\ldots, n; \ j=1,\ldots,p)$. 
Then  these covariates are evaluated at  $50$ equally spaced points $\{s_l\}_{l=1}^{50}$   
 between $0$ and $1$. 
We also generate  $n$ independent copies of  $q$ scalar  covariates $\bm{Z}_i = (Z_{i1}, \ldots, Z_{iq})^{\top} \in \mathbb{R}^q$ $(i=1,\ldots,n)$, 
based on the multivariate normal distribution with each component having mean $0$ and variance $1$, with correlations between the components   $\mbox{corr}(Z_{ij}, Z_{ik}) = 0.5^{|j-k|}$. 
We generate the outcomes   $Y_i$ $(i=1,\ldots,n)$ from: 
\begin{equation} \label{sim.model1}
\begin{aligned}
Y_i &=    \  \epsilon_i \ + \ 
\delta\bigg\{ \sum_{j=1}^8 \sin( \langle \eta_j, X_{ij} \rangle ) + \sum_{k=1}^8 \sin(Z_{ik}) \bigg\}
\ + \
 \\
&
4 (A_i -1.5)
\bigg[  \sin(\langle \beta_1, X_{i1} \rangle)  - \sin( \langle \beta_2, X_{i2} \rangle)  
+ \cos(Z_{i1})  - \cos(Z_{i2})  
+ \xi \left\{ \cos(  \langle X_{i1}, X_{i2} \rangle ) 
  + 
   \sin(Z_{i1} Z_{i2}) \right\}
 \bigg], 
\end{aligned}
\end{equation}
where the treatments  $A_i \in \{1,2\}$ are generated  with equal probability, 
independently of 
$(\bm{X}_i, \bm{Z}_i)$ 
and $\epsilon_i \sim \mathcal{N}(0,  0.5^2)$. 
In 
(\ref{sim.model1}),  there are  only four ``signal'' covariates  
 ($X_{i1}, X_{i2}, Z_{i1}$ and $Z_{i2}$) influencing the effect of 
 $A_i$ 
 on $Y_i$
  (i.e., $4$ treatment effect-modifiers).   The other $p + q -4$ covariates are ``noise'' covariates not critical in optimizing ITRs.   
We set $p=q=20$, therefore we consider a total of $40$ pretreatment covariates in this example. 
In (\ref{sim.model1}), 
 we set the single-index coefficient
  functions, $\beta_1$ and $\beta_2$,
   to be: 
 $\beta_{1}(s) = \bm{\Phi}(s)^\top \left(0.5, 0.5,  0.5,  0.5 \right)$ and 
 $\beta_{2}(s) = \bm{\Phi}(s)^\top \left(0.5, -0.5,  0.5,  -0.5 \right)$, respectively (see Figure~\ref{fig.rse}). 
 We set the coefficient functions 
$\eta_j$ $(j=1,\ldots,8)$ associated with the $X_{j}$  ``main''  effect to be: 
$\eta_j(s) = \bm{\Phi}(s)^\top \widetilde{\bm{\eta}}_j$, 
with each 
$\widetilde{\bm{\eta}}_j \in \mathbb{R}^4$ $(j=1,\ldots,8)$  following $\mathcal{N}(\bm{0}, \bm{I}_4)$  and  then rescaled to a unit $L^2$ norm $\lVert \widetilde{\bm{\eta}}_j \rVert =1$. 
The data model (\ref{sim.model1}) is indexed by a pair $(\delta, \xi)$. 
The parameter $\delta \in \{1,2\}$ controls the 
the contribution of the $(\bm{X}, \bm{Z})$ main effect component, $\delta \big\{\sum_{j=1}^{8}  \sin( \langle \eta_j, X_{ij} \rangle)  + \sum_{k=1}^{8} \sin(Z_{ik})\big\}$, to the variance of $Y$, 
in which $\delta = 1$ corresponds to a relatively  \textit{moderate} $(\bm{X}, \bm{Z})$ main  effect (about $4$ times greater than the interaction effect  when $\xi=0$) and $\delta = 2$ corresponds to a relatively \textit{large} $(\bm{X}, \bm{Z})$  main effect (about $16$ times greater than the interaction effect when $\xi=0$). In (\ref{sim.model1}), 
 the parameter  $\xi  \in \{0,1\}$ 
 determines whether 
 the $A$-by-$(\bm{X}, \bm{Z})$  interaction effect component 
  has an  additive regression structure $(\xi  = 0)$ of the form (\ref{the.model}) or whether 
  it deviates from an additive regression structure $(\xi  = 1)$. In the case of $\xi  =0$, the proposed CFAM (\ref{the.model}) is correctly specified, 
  whereas, for the case of  $\xi  =1$, it 
   is misspecified. 
For each simulation replication, we consider the following  
four approaches to estimating $\mathcal{D}^{opt}$\thad{:}

\begin{enumerate}
\item[1.]
The proposed 
 approach (\ref{LS5}),   
estimated via  Algorithm~\ref{algorithm1}, with 
the dimensions of the cubic $B$-spline basis 
for 
$\{g_j, h_k, \beta_j\}$  
set at 
$d_j = d_k =  m_j = 4 + (2n)^{1/5}$ (rounded to the closest integer), following  Corollary 3 of \cite{FAR}. 
 The sparsity tuning 
 parameter   $\lambda > 0$  is chosen to minimize $10$-fold cross-validated prediction error of the fitted models.

\item[2.]

The functional linear regression approach  of \cite{Ciarleglio.jrss.c}, 
 \begin{equation*} \label{the.mc.approach}
\underset{ \beta_j \in L^2[0,1], \alpha_k \in \mathbb{R} }{\text{minimize}} \  E
 \bigg\{ Y -   \sum_{j=1}^p  \langle \beta_{j}, X_{j} \rangle(A - 1.5 )  - \sum_{k=1}^q  \alpha_k Z_k  (A - 1.5 ) \bigg\}^2 
   + \lambda \bigg\{  \sum_{j=1}^p (\lVert \beta_j \rVert + 
\rho_j  \bm{\gamma}_j^\top \bm{S}_j  \bm{\gamma}_j)  + 
\sum_{k=1}^q |\alpha_k |) \bigg\}, 
\end{equation*}
which tends to result in a sparse set $\{ \beta_{j} \} \cup \{ \alpha_k \}$,  
performing estimation 
 based on representation (\ref{p.spline}) for the coefficient functions $\beta_j$ given $m_j = 10$,  
  with an associated $m_j \times m_j$ $P$-spline penalty matrix $(\bm{S}_j)$  
  to ensure appropriate smoothness. 
The tuning parameters  $\lambda >0$ and $\rho = \rho_j >0$  $(j=1,\ldots,p)$ are chosen to minimize a $10$-fold cross-validated prediction error \citep{Ciarleglio.jrss.c}, and 
the ITR is given by: $\widehat{\mathcal{D}}^{opt}(\bm{X}, \bm{Z}) =  \operatorname*{arg\,max}_{a \in \{1,\ldots,L \} } \big\{  \sum_{j=1}^p (a - 1.5 ) \langle \widehat{\beta}_j,  X_{j} \rangle +  (a - 1.5 )  \sum_{k=1}^q \widehat{\alpha}_{k} Z_k \big\}$. 
Since  the component functions $\{g_j, h_k \}$ associated with \cite{Ciarleglio.jrss.c} are restricted to be linear
\big(i.e., we restrict them to 
$g_{j}(\langle \beta_j,  X_{j} \rangle, A) =   \langle \beta_j,  X_{j} \rangle(A-1.5)$ 
and 
$h_{k}(Z_k,A) = \alpha_{k} Z_k (A-1.5)$\big) corresponding to a special case of  CFAM,  
we call the model of \cite{Ciarleglio.jrss.c}, a CFAM with \textit{linear} component functions (CFAM-lin) 
for the notational simplicity.

\item[3.]
The outcome weighted learning  \citep[OWL;][]{Zhao.2012} method based on a linear kernel (OWL-lin), implemented in the R-package \texttt{DTRlearn}.  
 Since there is  no currently available OWL method that deals with functional covariates, we compute a scalar summary of each functional covariate, i.e., $\bar{X}_{j}  = \int_{0}^1 X_{j}(s)ds \in \mathbb{R}$,  
and use $\bar{X}_{j}$ along with the other scalar covariates $Z_{k}$ as inputs to the augmented (residualized)  OWL procedure. 
To improve its efficiency,  we employ  the augmented OWL 
approach of \cite{Augmented.OWL},  
which amounts to pre-fitting a linear model for $\mu$ in (\ref{the.model}) via Lasso \citep{Tibshirani.lasso} and residualizing the response $Y$. 
The tuning parameter $\kappa$ in \cite{Zhao.2012} is chosen from the grid of $(0.25, 0.5, 1, 2, 4)$ (the default setting of \texttt{DTRlearn}) based on a $10$-fold cross-validation.

\item[4.]
The same approach as in 3 but based on a Gaussian  radial basis function kernel (OWL-Gauss) in place of a linear kernel. 
The inverse bandwidth parameter  $\sigma_n^2$ in \cite{Zhao.2012} is chosen from the grid of $(0.01, 0.02, 0.04, \ldots, 0.64, 1.28)$ 
and $\kappa$ is chosen from the grid of $(0.25, 0.5, 1, 2, 4)$, based on a $10$-fold cross-validation. 

\end{enumerate}

Throughout the paper, for CFAM and CFAM-lin, 
we fit the $(\bm{X},\bm{Z})$ ``main'' effect on $Y$ based on the (misspecified) linear model with the na\"{\i}ve scalar averages of $X_j$,  i.e., 
 $\bar{X}_{j}$, along with $Z_k$, fitted via Lasso with 10-fold cross-validation for the sparsity parameter 
 and utilize the ``residualized'' response  $Y - \widehat{\mu}(\bm{X}, \bm{Z})$.  
 For each simulation run,   
 we estimate $\mathcal{D}^{opt}$ from each of the above four methods 
 based on a training set (of size $n \in \{250, 500\}$), and  to evaluate these methods, we 
compute  the value 
$V(\widehat{\mathcal{D}}^{opt}) = E[ E[ Y |\bm{X}, \bm{Z}, A= \widehat{\mathcal{D}}^{opt}(\bm{X}, \bm{Z})] ]$ 
given each estimate  $\widehat{\mathcal{D}}^{opt}$,    
using a Monte Carlo approximation 
based on a separate random sample of size $10^3$. 
Since we know the true data generating model in simulation studies, 
the optimal 
 $\mathcal{D}^{opt}$  can be determined 
for each simulation run. 
Given each estimate $\widehat{\mathcal{D}}^{opt}$ of $\mathcal{D}^{opt}$, 
we report  $V(\widehat{\mathcal{D}}^{opt})  - V(\mathcal{D}^{opt})$,  
as the performance measure of $\widehat{\mathcal{D}}^{opt}$. 
A larger (i.e., less negative) value of the measure indicates  better performance.

\begin{figure} [H]
\begin{center}
\begin{tabular}{c}
\includegraphics[width=6.4in, height = 1.8 in]{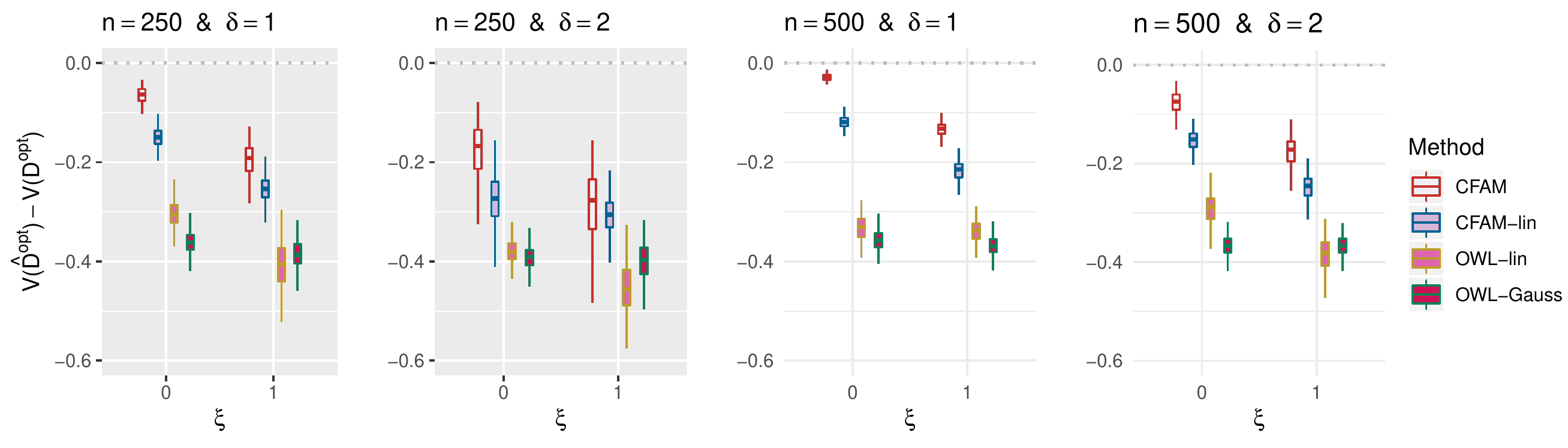}
\end{tabular}
\end{center}
\vspace{-0.15in}
\caption{
Boxplots obtained from  $200$ Monte Carlo simulations 
comparing $4$ approaches to estimating $\mathcal{D}^{opt}$, 
given each scenario indexed by $\xi \in \{0,1\}$, $\delta \in \{1,2\}$ and  $n\in \{250, 500\}$. 
The dotted horizontal line represents the optimal value corresponding to $\mathcal{D}^{opt}$.   
} \label{fig.sim.result2}
\end{figure}

In Figure \ref{fig.sim.result2},  
we present boxplots, obtained from $200$ simulation runs, 
 of the normalized
  values $V(\widehat{\mathcal{D}}^{opt})$ 
 (normalized by the optimal values $V(\mathcal{D}^{opt})$)  of the decision rules  $\widehat{\mathcal{D}}^{opt}$ 
estimated from the four approaches, 
for each combination of $n \in \{250, 500 \}$, 
$\xi \in \{0, 1\}$ (corresponding to   \textit{correctly-specified} or \textit{mis-specified} CFAM interaction models, respectively) and 
 $\delta \in \{1, 2 \}$  (corresponding to \textit{moderate} or  \textit{large} main effects, respectively). 
The results in Figure~\ref{fig.sim.result2} 
indicate that the proposed method (CFAM)  
 outperforms all other approaches.  
 In particular, 
  if the sample size is relatively large $(n=500)$, 
for a correctly-specified CFAM  ($\xi = 0$) interaction underlying model, 
the proposed method gives a close-to-optimal performance in comparison to  $\mathcal{D}^{opt}$. 
With nonlinearities present in the underlying model (\ref{sim.model1}), 
 CFAM-lin, which assumes a stringent linear structure on the interaction effect term, 
is outperformed by  
CFAM that utilizes the flexible component functions  $g_{j}(\cdot,a)$ and $h_{k}(\cdot,a)$, while 
substantially outperforming the OWL-based approaches. 

In Section A.5 of Supporting Information, 
we have also considered a similar set of experiments under a ``\textit{linear}''  $A$-by-$\bm{X}$ interaction effect, 
 in which CFAM-lin 
outperforms CFAM, but by a relatively small amount, 
whereas 
 if the underlying model deviates from the exact linear structure and $n=500$, 
 CFAM tends to outperform CFAM-lin.
 This suggests that, in the absence of prior knowledge about the form of the interaction effect, 
the more flexible 
 CFAM 
 that accommodates nonlinear treatment effect-modifications
 can be set as a default approach over CFAM-lin for optimizing ITRs. The estimated values of the OWL methods using linear and Gaussian kernels, respectively, 
 are similar to each other, but both are outperformed by CFAM, even when CFAM is incorrectly specified (i.e., when $\xi = 1$), 
 as the current OWL methods do not directly deal with the functional pretreatment covariates. 
When the $(\bm{X},\bm{Z})$  ``main'' effect dominates the $A$-by-$(\bm{X},\bm{Z})$ interaction effect 
 (i.e., when $\delta = 2$), 
although the increased magnitude of this nuisance effect dampens the performance of all approaches to estimating $\mathcal{D}^{opt}$,  
the proposed approach outperforms all other methods.

In Table~\ref{t:one}, we additionally illustrate the estimation performance for model parameters $\beta_1$ and $\beta_2$ when $\xi = 0$ (i.e., when CFAM is correctly specified) with varying $\delta \in \{1,2\}$ and $n \in \{250, 500, 1000 \}$, based on the root squared error $\mbox{RSE}(\beta_j) = \sqrt{ \int (\widehat{\beta}_j(s) - \beta_j(s))^2 ds }$ $(j=1,2)$.  
In Figure~\ref{fig.rse}, we display typical  CFAM 
 estimates $\widehat{\beta}_j$ of $\beta_j$ from 10 random samples, for 
each sample size $n$ (for the case of $\delta = 1$). 
With sample size increasing, the estimators $\widehat{\beta}_j$  get  close to the true coefficient functions $\beta_j$.  

\begin{table}
\caption{The root squared error (RSE) of the estimates $\widehat{\beta}_j$ for $\beta_j$ $(j=1,2)$ for  varying sample size $n \in \{250, 500, 1000\}$, 
when the ``main'' effect of $(\bm{X}, \bm{Z})$ is moderate $(\delta = 1)$   and large $(\delta = 2)$, respectively.  } \label{t:one}
\begin{center}
\begin{tabular}{lrrrrrr}
\hline
	 & \multicolumn{3}{c}{$\delta =1$ (\textit{Moderate} ``main'' effect)}  &   \multicolumn{3}{c}{$\delta=2$ (\textit{Large} ``main'' effect)} \\ \hline
\quad	\quad \quad n & \multicolumn{1}{c}{$250$} &  \multicolumn{1}{c}{$500$} & \multicolumn{1}{c}{$1000$} &   \multicolumn{1}{c}{$250$} &  \multicolumn{1}{c}{$500$} & \multicolumn{1}{c}{$1000$} \\ \hline
 $\mbox{RSE}(\beta_1)$ & 0.53(0.08) & 0.34(0.02) & 0.26(0.02) & 0.60(0.14) & 0.38(0.05)  &  0.29(0.03)   \\
 $\mbox{RSE}(\beta_2)$ & 0.53(0.06) & 0.34(0.02) & 0.27(0.01)   &  0.59(0.13) & 0.39(0.07)  & 0.29(0.03)  \\
\hline
\end{tabular}
\end{center}
\end{table}

\begin{figure} [H]
\begin{center}
\begin{tabular}{c}
\includegraphics[width=5.8in, height = 3 in]{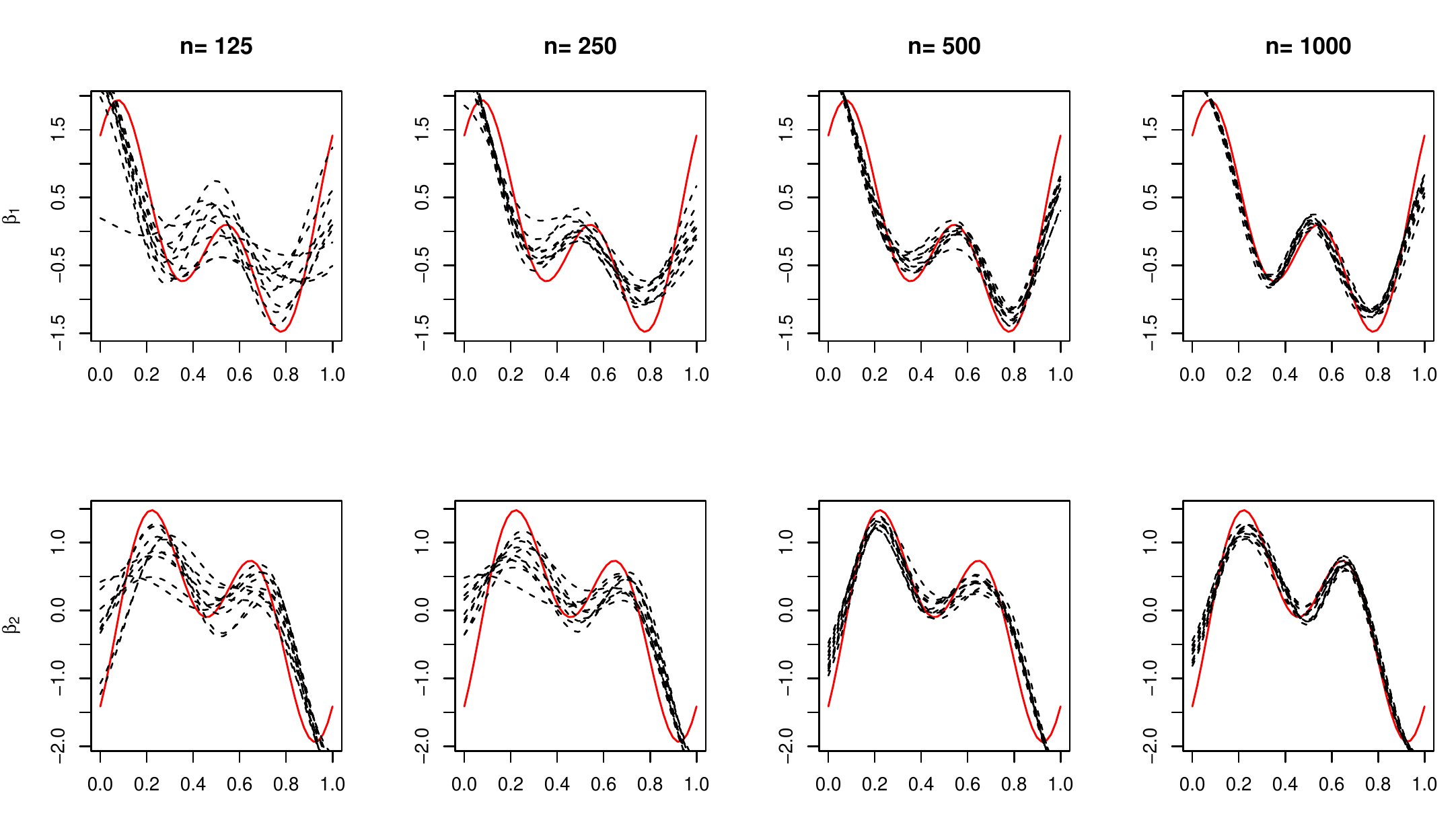}
\end{tabular}
\end{center}
\vspace{-0.2in}
\caption{
An illustration of typical $10$ CFAM sample 
 estimates $\widehat{\beta}_j(s)$ (black dashed curves) for the parameters 
$\beta_j(s)$ (the red solid curves), for $j=1$ and $2$ in the top and bottom panels,  respectively, 
with a varying training 
 sample size $n \in \{125, 250, 500, 1000\}$ for the case of $\delta = 1$. 
} \label{fig.rse}
\end{figure}

\subsection{Treatment effect-modifier variable selection performance} \label{VS.sim}

In this subsection, we will report simulation results for 
the treatment effect-modifier selection 
among $\{X_j, j=1,\ldots,p\} \cup \{Z_k, k=1,\ldots,q \}$. 
The complexity of the $(\bm{X}, \bm{Z})$-by-$A$ interaction terms of  CFAM (\ref{the.model}) 
can be summarized in terms of the size (cardinality) 
 of the index set of 
$\{g_j, j=1,\ldots,p\} \cup \{h_k, k=1,\ldots,q\}$ 
that are not identically zero, 
each of which can be either correctly or incorrectly estimated to be equal to zero. 
As in Section~\ref{ITR.sim}, 
we generate $200$ datasets  
based on (\ref{sim.model1}),  
with varying  $\xi \in \{0,1\}$, $\delta \in \{1,2\}$ and  sample size 
$n \in \{50, 100, 200, \ldots, 700, 800\}$ and 
 $p=q=20$, 
i.e., we consider a total of $p+q=40$ potential treatment effect-modifiers, 
among which there are only $4$ ``true'' treatment effect-modifiers. 

\begin{figure} [H] 
\begin{center}
\begin{tabular}{c} 
\includegraphics[width=6.4in, height = 4.0in]{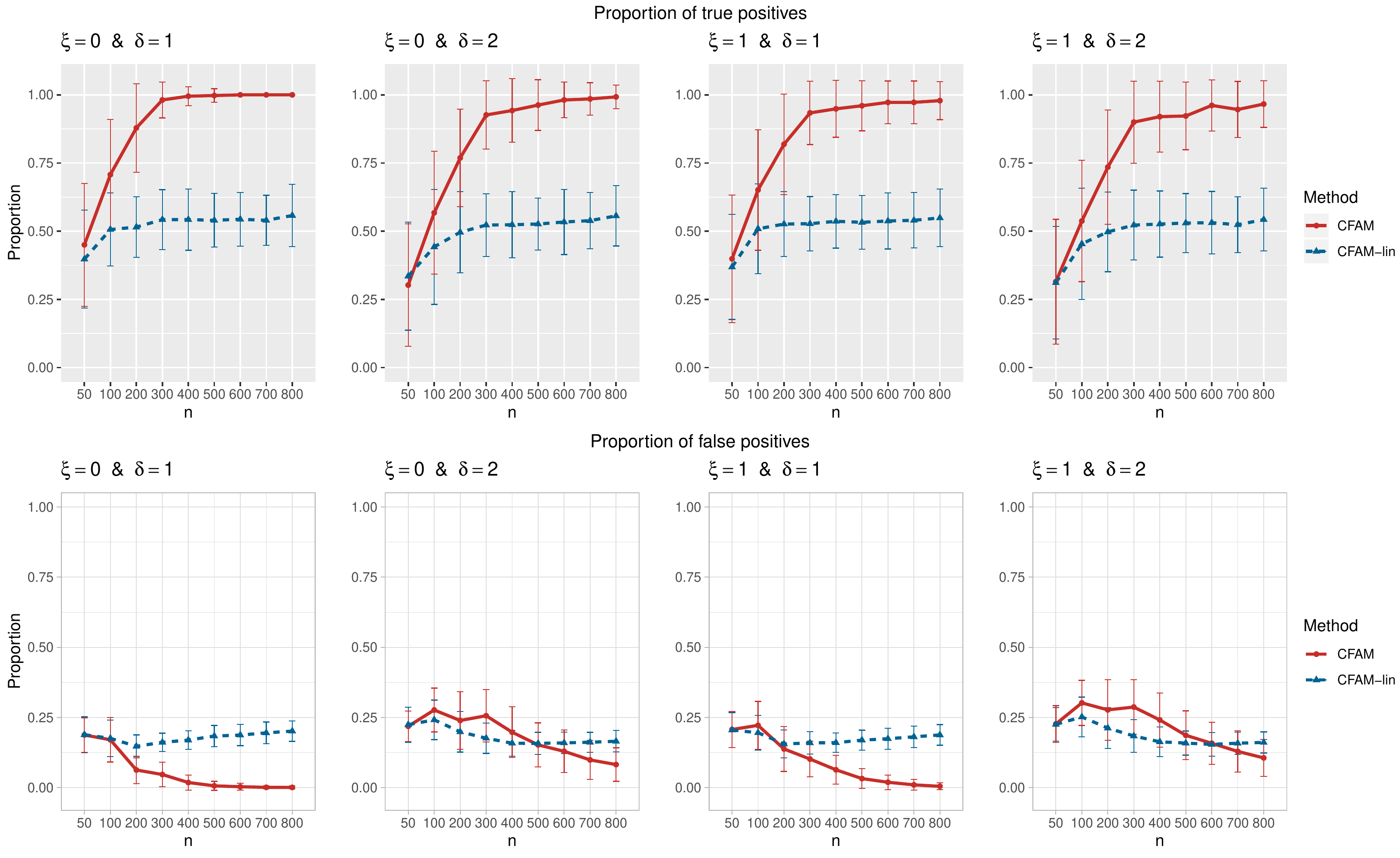}
 \end{tabular}
\end{center}
\vspace{-0.15in}
\caption{The proportion of the relevant covariates (i.e., the treatment effect-modifiers) 
correctly selected (the ``true positives''; the top gray panels),
 and the ``noise'' covariates  incorrectly selected (the ``false positives''; the bottom white panels),  respectively (and $\pm 1$ standard deviation), 
 with a varying sample size $n \in \{50, 100, 200, \ldots, 800\}$,  
 for each combination of $\xi \in \{0,1\}$ and $\delta \in \{1,2\}$. 
} \label{fig.sim.result.vs}
\end{figure}

Figure~\ref{fig.sim.result.vs} summarizes the results of the treatment effect-modifier covariate selection performance 
with respect to 
the true/false positive rates (the top/bottom panels, respectively), comparing 
the proposed CFAM  and  the CFAM-lin of \cite{Ciarleglio.jrss.c}. 
The results are reported as the averages (and $\pm 1$ standard deviations) across the $200$ simulated datasets, for each simulation scenario. 
 Figure~\ref{fig.sim.result.vs} illustrates that 
the proportion of \textit{correct} selection 
 out of the $4$ \textit{true} treatment effect-modifiers 
(i.e., the ``true positive'' rate; the top gray panels)
 of CFAM (the red solid curves) tends to $1$, as $n$ increases  from $n=50$ to $n=800$, 
whereas the proportion of \textit{incorrect} selection 
(i.e., the ``false positive'' rate; the bottom white panels) 
out of the $36$  \textit{irrelevant} ``noise'' covariates  
 tends to $0$; 
 these proportions tend to either $1$ or $0$ quickly 
  for \textit{moderate} main effect ($\delta = 1$) scenarios compared to \textit{large} main effect  ($\delta = 2$) scenarios.   
On the other hand,  
the proportion of correct  selections 
 for CFAM-lin (the blue dotted curves), even with a large $n$, 
tends to be only around $0.55$, due to the stringent linear model assumption on the from of the $(\bm{X},\bm{Z})$-by-$A$ interaction effect.

\section{Application}\label{sec.application}

In this section, we illustrate the utility of CFAM for optimizing ITRs, using data from an RCT \citep[][]{embarc} comparing an antidepressant and placebo for treating major depressive
disorder. 
The study collected various scalar and functional patient characteristics at baseline, including electroencephalogram (EEG) data. 
Study participants were randomized to either  placebo  ($A=1$) or an antidepressant (sertraline) ($A=2$). Subjects were monitored 
for 8 weeks after initiation of treatment. The primary endpoint of interest was the Hamilton Rating Scale for Depression (HRSD)  score at week 8. 
The outcome $Y$ was taken  to be the improvement in symptoms severity from baseline to week $8$ taken as the difference: 
week 0 HRSD score 
- week 8 HRSD score (larger values of the outcome $Y$ are considered desirable).  

\begin{figure} 
\begin{center}
  \includegraphics[width=2.5in, height = 2.2in]{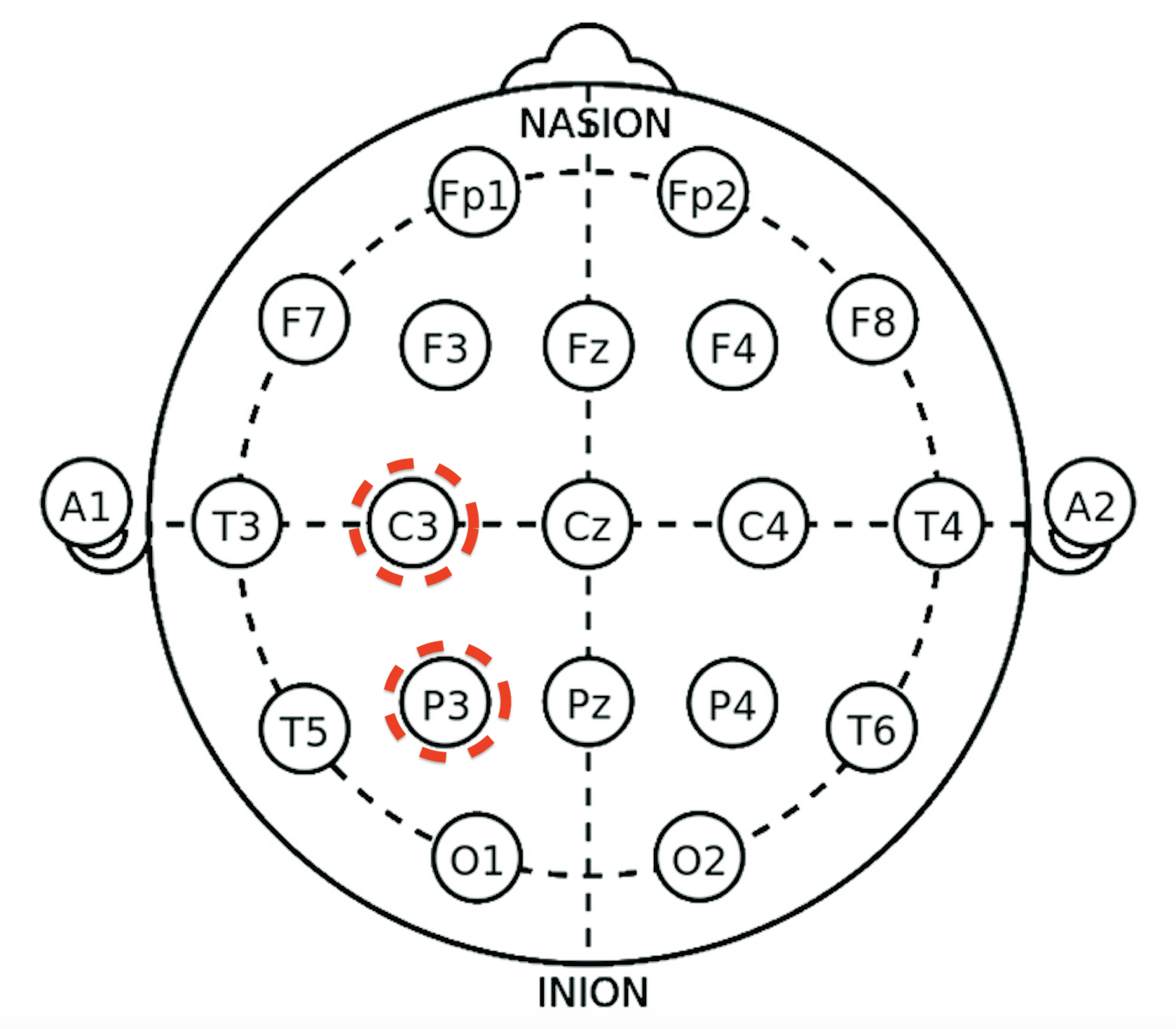}   \\
\begin{tabular}{c}  \includegraphics[width=6.5in, height = 4.1in]{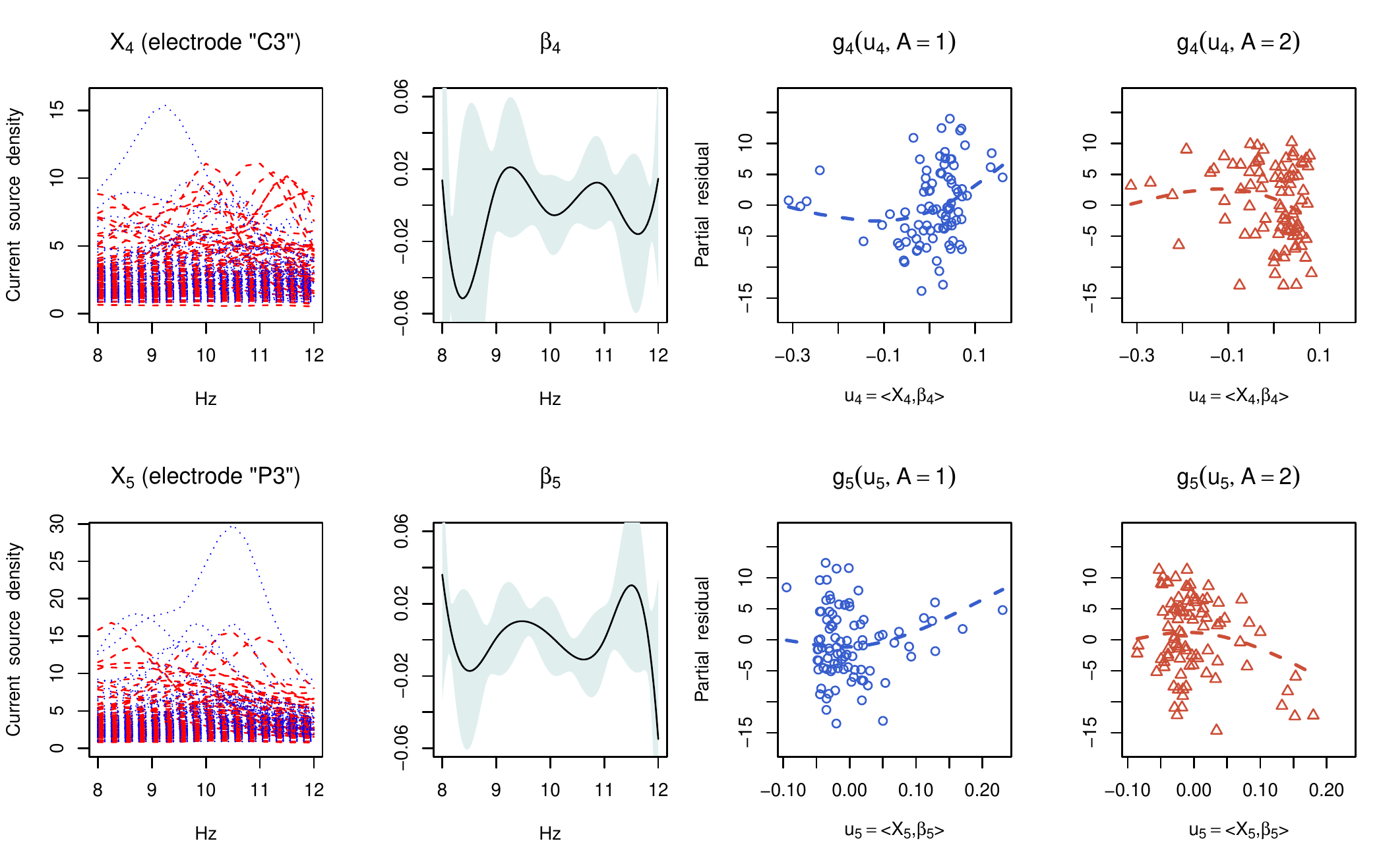} \end{tabular} 
\end{center}
\vspace{0.1in}
\caption{
{\bf Top}: The locations for the $19$ electrode channels (``A1'' and ``A2'' were not used). 
Those marked in red circles are the  selected electrodes from the proposed approach:  ``C3'' and ``P3''.
{\bf Bottom}: First column: 
observed current source density (CSD) curves from the selected channels $X_{4}$ (``C3'') and $X_{5}$ (``P3''),  
over the Alpha band ($8$ to $12$ Hz). 
Second column: 
the estimated single-index coefficient functions ($\beta_{4}$ and $\beta_{5}$) for the selected channels $X_{4}$ and $X_{5}$ (and the associated $95 \%$ confidence bands, conditioning on the $j$th partial residual and the $j$th component function $\widehat{g}_j$).  
Third and fourth columns: 
the scatter plots of the ($j$th; $j=4, 5$, top and bottom, respectively) partial residuals vs. the estimated single-indices  $u_4 = \langle X_{4}, \beta_{4} \rangle$ 
and $u_5 =\langle X_{5}, \beta_{5} \rangle$, respectively, 
for the placebo  $A=1$ (third column, blue circles) and sertraline $A=2$ (fourth column, red  triangles) treated individuals, 
with the estimated treatment-specific component functions $g_{j}(u_j, A)$ $(A=1,2)$ $(j = 4, 5)$ (the dashed curves) overlaid. } 
\label{fig.embarc.1}
\end{figure}

There were $n= 180$ subjects. 
We considered  $p=19$ pretreatment functional covariates consisting of
the current source density (CSD) amplitude spectrum curves over the Alpha frequency range (observed while the participants' eyes were open),  
measured from a subset of EEG channels from a total of 72 EEG electrodes   
which gives a fairly good spatial coverage of the scalp. 
The locations for these $19$ electrodes are  indicated in the top panel of Figure~\ref{fig.embarc.1}. 
The Alpha frequency band ($8$ to $12$ Hz) considered as a potential biomarker of antidepressant response   \citep[e.g.,][]{Wade2016} 
was scaled to $[0, 1]$, hence each of the functional covariates $X = ( X_1(s), \ldots, X_{19}(s))$ was defined on the interval $[0, 1]$. 
We also considered  $q=5$ baseline scalar covariates consisting of  the week 0 HRSD score ($Z_{1}$),  sex ($Z_{2}$), 
 age at evaluation ($Z_{3}$), 
 word fluency ($Z_4$) and Flanker accuracy ($Z_5$) cognitive test scores,  which were identified as predictors of differential treatment response in a previous study \citep{SAM2020}.  
In this dataset, 49\% of the subjects were randomized to the sertraline $(A=2)$.  
The average outcomes $Y$ for the sertraline and placebo groups were 
7.41 and 6.29, respectively.  The means (and standard deviations) of $Z_1$, $Z_3$, $Z_4$ and $Z_5$ were 
18.59 (4.44), 37.7 (13.57), 38 (11.42) and 0.19 (0.11), respectively, and 67\% of the subjects were female.

The proposed  CFAM approach  (\ref{LS5}) selected two functional covariates: 
``C3'' ($X_{4}$) and ``P3'' ($X_{5}$)  (the selected electrodes are indicated by the red dashed circles in the top panel of Figure~\ref{fig.embarc.1}), 
 and a scalar covariate: ``Flanker accuracy test'' ($Z_{5}$). 
In the first column of  Figure~\ref{fig.embarc.1}, we display CSD curves  
corresponding to the selected two  functional covariates,  
$X_4(s)$ and $X_5(s)$, 
 from the $180$ subjects. 
In the second column  of  Figure~\ref{fig.embarc.1}, we display the estimated  coefficient functions, $\widehat{\beta}_4(s)$ and $\widehat{\beta}_5(s)$ 
(with $95 \%$ confidence bands conditional on the $j$th partial residual and the $j$th component function $\widehat{g}_j$),  
associated with those selected covariates.

The coefficient functions  $\widehat{\beta}_j(s)$ summarizing the $X_j(s)$ 
lead to data-driven \textit{indices} $u_j = \langle \widehat{\beta}_j, X_j \rangle \in \mathbb{R}$ that are linked to differential treatment response by 
two estimated nonzero component functions, $\widehat{g}_j(u_j, A)$ $(j=4, 5)$ in this example. 
In Figure~\ref{fig.embarc.1}, 
the fitted component functions,  $\widehat{g}_{j}(u_j, A)$ 
associated with the placebo $A=1$ (in the third column) 
and the active drug $A=2$ (in the fourth column),    
are displayed, along with the corresponding partial residuals. 
Roughly put, in Figure~\ref{fig.embarc.1}, the  placebo $(A=1)$ 
effect tends to increase with 
 the index $u_j$ $(j=4,5)$ 
whereas the sertraline $(A=2)$ effect decreases with the index. 
In the second column of Figure~\ref{fig.embarc.1}, both $\beta_4$ and $\beta_5$ put a  bulk of their negative weight on lower frequencies (8 to 9 Hz), meaning that 
patients whose CSD values are small in those frequency regions would have large values of $\langle \beta_j, X_j \rangle$,
 over the values which the  placebo effects are predicted to be relatively strong, in comparison to the sertraline effects.

\begin{figure} 
\begin{center}
\begin{tabular}{c}  
\includegraphics[width=4.1in, height = 2.1in]{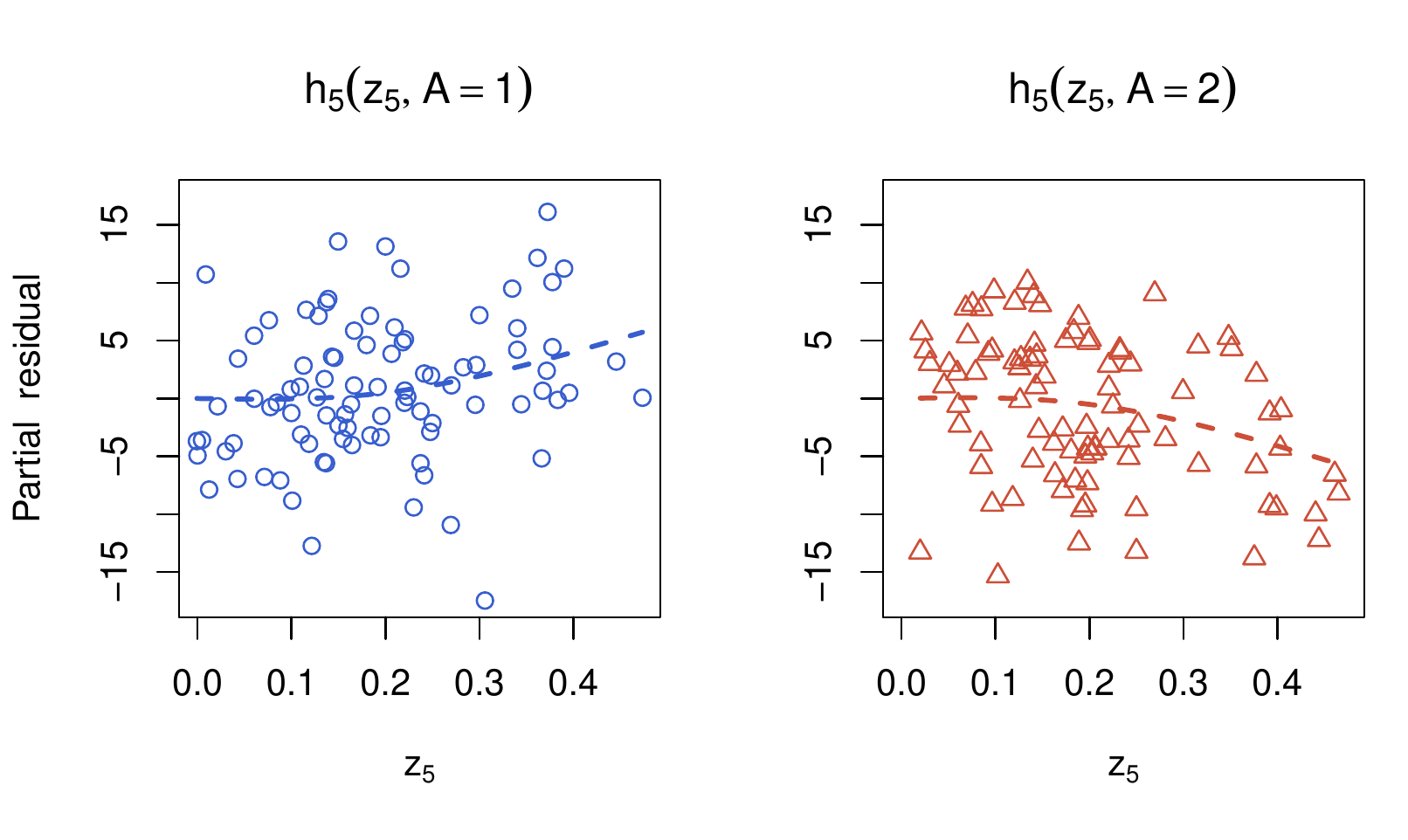}
\end{tabular}  
\end{center}
\vspace{0.1in}
\caption{
The scatter plots of the $k$th partial residual vs. the $k$th  scalar covariate, 
for the selected $5$th scalar covariate $Z_5$  
 ``Flanker accuracy test score,''  
for the placebo  $A=1$ (first column, blue circles)
 and sertraline $A=2$ (second column, red triangles) 
treated individuals, 
with the estimated treatment-specific component functions $h_{5}(z_5, A)$ $(A=1,2)$ (the dashed curves) overlaid. 
} \label{fig.embarc.2}
\end{figure}

\begin{figure} 
\begin{center}
\begin{tabular}{c}
 \includegraphics[width=4.2in, height = 2.4 in]{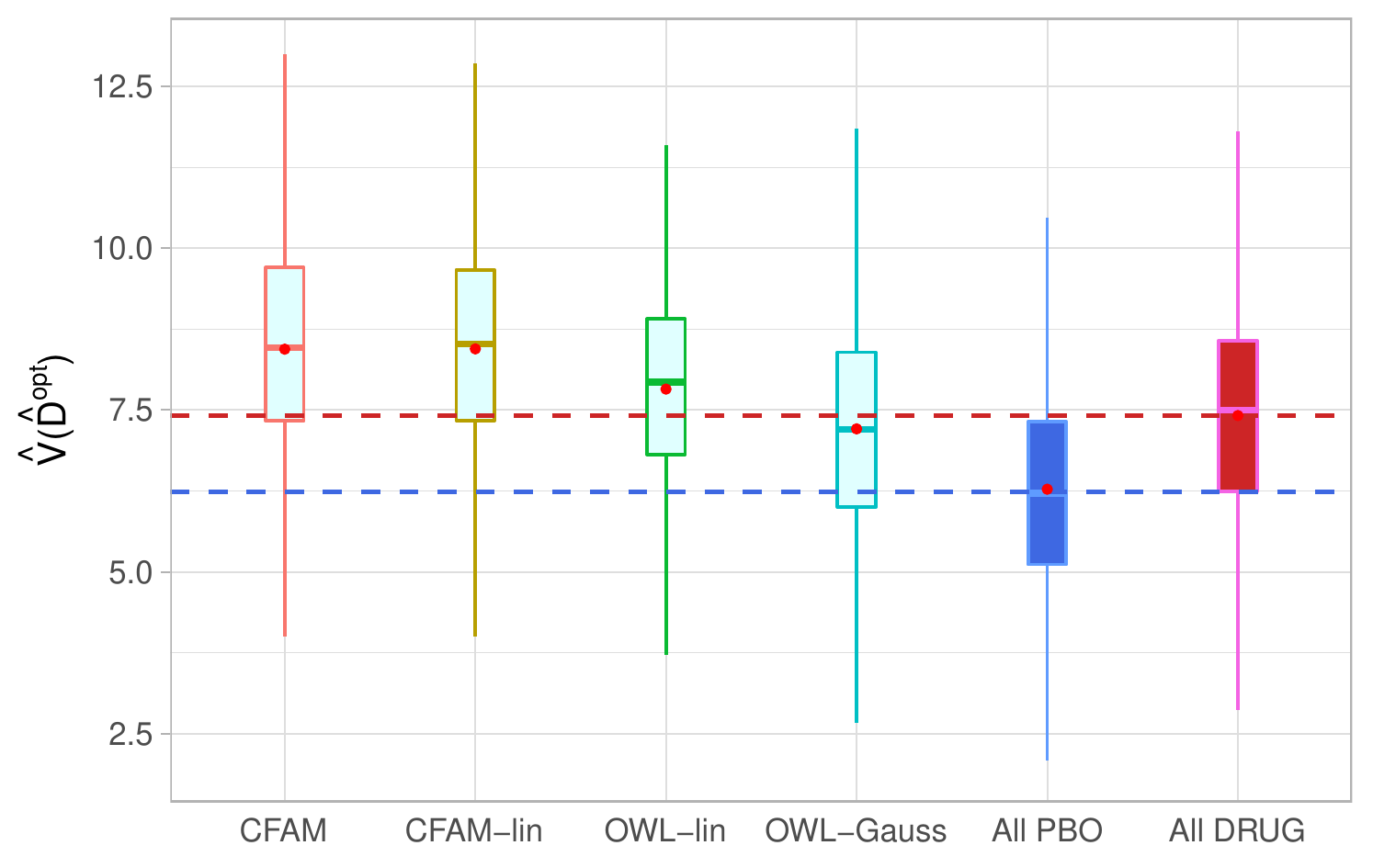}
\end{tabular}
\end{center}
\vspace{0.1in}
\caption{
Boxplots of the estimated values of the treatment rules $\widehat{\mathcal{D}}^{opt}$ estimated from $6$ approaches, obtained from $500$ randomly split testing sets. Higher values are preferred. } \label{fig.embarc.4}
\end{figure}

To evaluate the performance of ITRs  $(\widehat{\mathcal{D}}^{opt})$  estimated from  the four different approaches 
described in Section~\ref{sec.simulation}, we randomly split the data into a training set and a testing set  (of size $\widetilde{n}$) using a ratio of 
$5:1$, 
 replicated $500$ times, 
 each time estimating an ITR $\widehat{\mathcal{D}}^{opt}$ 
 based on the training set, and its ``value'' $V(\widehat{\mathcal{D}}^{opt})  = E[ E[ Y |\bm{X}, \bm{Z}, A= \widehat{\mathcal{D}}^{opt}(\bm{X}, \bm{Z})]]$, by an inverse probability weighted estimator \citep{Murphy2005}
$\widehat{V}(\widehat{\mathcal{D}}^{opt}) = \sum_{i=1}^{\widetilde{n}} Y_{i} I_{(A_i = \widehat{\mathcal{D}}^{opt}(\bm{X}_i, \bm{Z}_i) )} / \sum_{i=1}^{\widetilde{n}} I_{(A_i =\widehat{\mathcal{D}}^{opt}(\bm{X}_i, \bm{Z}_i))}$, 
computed   
based on the testing set (of size $\widetilde{n}$). 
  For comparison, 
we also include two na\"{\i}ve rules: 
 treating all patients with placebo (``All PBO'')  and treating all patients with the active drug (``All DRUG''), each regardless of the individual patient's characteristics $
 (\bm{X},\bm{Z})$.   The resulting boxplots obtained from the $500$ random splits are illustrated in Figure~\ref{fig.embarc.4}.

The results in Figure~\ref{fig.embarc.4} 
demostrate that 
CFAM and CFAM-lin perform at a similar level, 
showing a  clear advantage over the both OWL-lin and OWL-Gauss, 
suggesting that 
 regression utilizing the functional nature of the EEG measurements, 
that 
 targets  the treatment-by-functional covariates interactions 
 is well-suited in this example. 
Specifically, in Figure~\ref{fig.embarc.4}, the superiority of CFAM (or CFAM-lin) over 
the policy of 
treating everyone with 
the drug 
 (All DRUG)  was of similar magnitude of the superiority of  All DRUG over All PBOs. This suggests that accounting for patient characteristics can help treatment decisions.  
In this example, 
as can be observed from the third and fourth columns of Figures~\ref{fig.embarc.1}  
and \ref{fig.embarc.2}, 
the estimated nonlinear  treatment effect-modification is rather modest. As a result,
 the performances of CFAM and CFAM-lin are comparable to each other. 
However, as demonstrated in Section~\ref{sec.simulation}, 
the more flexible CFAM can be employed as a default approach over  
CFAM-lin, 
 allowing for 
 potentially essential nonlinearities in treatment effect-modification.

\section{Discussion} \label{sec.conclusion}

We have developed a functional additive 
regression approach specifically focused on extracting possibly nonlinear pertinent interaction effects 
between treatment and multiple functional/scalar covariates, which is of paramount importance in developing effective ITRs for precision medicine.
This is accomplished by imposing   appropriate structural constraints, performing treatment effect-modifier selection 
and extracting one-dimensional functional indices. 
The estimation 
approach 
utilizes
  an efficient coordinate-descent for the component functions 
and a functional linear model estimation procedure  for the  coefficient functions. 
The proposed functional regression for ITRs
 extends existing 
methods by incorporating possibly nonlinear treatment-by-functional covariates interactions. 
 Encouraged by our simulation results and the application,  future work will investigate the asymptotic properties of the method related to variable selection and estimation consistency, 
 and  a hypothesis testing framework for  significant interactions between treatment and functional covariates.

\begin{center}
{\large\bf SUPPLEMENTARY MATERIAL}
\end{center} 
Supplementary Material  at the end of the document provides additional technical details referred to in the main paper, 
including the proof of Theorem~\ref{theorem1}. 
Supplementary Material also presents additional simulation examples, including a set of simulation experiments with a 
 ``linear'' $A$-by-$(\bm{X},\bm{Z})$ interaction effect scenario.
 
\begin{description}
\item[R-package:]  \texttt{R}-package \texttt{famTEMsel} (Functional Additive Models for Treatment Effect-Modifier Selection) contains \texttt{R}-codes to perform the methods proposed in the article, and is publicly available on \texttt{GitHub} (\texttt{syhyunpark/famTEMsel}). 
\end{description}

\section*{Acknowledgments}

This work was supported by National Institute of Health (NIH) grant 5 R01 MH099003.

\section*{Conflict of interest}

None declared.

\bibliographystyle{biom}
\bibliography{refs}

\newpage

\appendix

\section*{SUPPLEMENTARY MATERIAL}

\renewcommand{\theequation}{S.\arabic{equation}}
\renewcommand{\thefigure}{S.\arabic{figure}}
\renewcommand{\thetable}{S.\arabic{table}}
\renewcommand{\bibnumfmt}[1]{[S.#1]}
\renewcommand{\citenumfont}[1]{S.#1}

\setcounter{equation}{0}
\setcounter{figure}{0}
\setcounter{table}{0}

\section*{Appendix A: Technical details and additional simulations} 
\label{s:appendixA}

\subsection*{A.1. Description of the constrained least squares criterion in Section 2.1}

In Section 2 of the main manuscript, we introduce the constrained functional  additive model (CFAM) for the $(\bm{X},\bm{Z})$-by-$A$ interaction effect: 
\begin{equation} \label{the.model}
\begin{aligned}
Y \ = 
\mu(\bm{X}, \bm{Z}) 
+ 
\sum_{j=1}^p g_{j}(\langle X_j, \beta_{j} \rangle, A ) 
+ 
\sum_{k=1}^q h_{k}(Z_k, A) \  + \  \epsilon, 
\end{aligned} 
\end{equation}
with $\beta_j \in \Theta$, 
subject to the constraint on the component functions $g_j \in \mathcal{H}_j^{(\beta_j)}$ $(j=1,\ldots,p)$ and $h_k \in \mathcal{H}_k$ $(k=1,\ldots,q)$: 
 \begin{equation}   \label{the.condition}
\begin{aligned}
E[g_{j}( \langle X_j, \beta_j \rangle, A) | X_j ]  &
 = 0 \quad  
  \mbox{(almost surely)} \quad  
   (\forall \beta_j \in \Theta)
   \quad (j=1,\ldots,p) \quad \mbox{and} \\
E[h_{k}( Z_k, A) | Z_k ]   & 
= 0 \quad  
  \mbox{(almost surely)} \quad (k=1,\ldots,q),  
  \end{aligned}
\end{equation}
in which the expectation is taken with respect to the distribution of $A$ given $X_j$ (or $Z_k$),   
and $\epsilon \in \mathbb{R}$ is a mean zero noise with finite variance,  
and the form of the squared integrable functional $\mu$ in (\ref{the.model}) is left unspecified. 

Under model (\ref{the.model}) subject to (\ref{the.condition}),  
the ``true'' (i.e., optimal) 
 functional components, which we denote by 
$\{g_j^\ast, j=1,\ldots,p\} \cup \{\beta_j^\ast, j=1,\ldots,p\} \cup \{ h_k^\ast, k=1,\ldots,q\}$ 
 that
  constitute the $(\bm{X},\bm{Z})$-by-$A$ interaction effect, can be specified and viewed as the solution to the following constrained least squares problem:  
 \begin{equation} \label{the.criterion}
\begin{aligned}
\{ g_{j}^\ast, \beta_{j}^\ast, h_{k}^\ast  \} 
\quad = \quad & \underset{ g_{j}   \in \mathcal{H}_j^{(\beta_j)}, \beta_j \in \Theta,  h_k \in \mathcal{H}_k }{\text{argmin}}
& &E \bigg[ \bigg\{ Y -  \mu( \bm{X},\bm{Z})  -  \sum_{j=1}^p g_{j,A}(\langle X_j, \beta_j \rangle)  
- \sum_{k=1}^q h_{k,A}(Z_k) 
   \bigg\}^2 \bigg]  \\
& \text{subject to} & & E[g_{j,A}(\langle X_j, \beta_j \rangle) | X_j ]  = 0 \quad 
\forall \beta_j \in \Theta \quad 
(j=1,\ldots,p) \quad \mbox{and}  \\
& 			    & & E[h_{k,A}( Z_k) | Z_k ]  = 0 \quad (k=1,\ldots,q),  
\end{aligned}
\end{equation}
in which $\mu(\bm{X}, \bm{Z})$ is 
the ``main'' effect component  appeared in 
model (\ref{the.model}) (and is considered as fixed in (\ref{the.criterion})). 
In particular, on the right-hand side of (\ref{the.criterion}), 
the expected squared error criterion term can be expanded as: 
\begin{equation*}\label{the.criterion2}
\begin{aligned}
& 
 \argmin_{ g_{j}   \in \mathcal{H}_j^{(\beta_j)}, \beta_j \in \Theta,  h_k \in \mathcal{H}_k  }  
\mathbb{E}\bigg[  
\bigg\{ Y -  \sum_{j=1}^p g_{j,A}(\langle X_j, \beta_j \rangle) - \sum_{k=1}^q h_{k,A}(Z_k)  \bigg\}^2   
+ 2  \mu(\bm{X}, \bm{Z}) \bigg\{\sum_{j=1}^p g_{j,A}(\langle X_j, \beta_j \rangle) + \sum_{k=1}^q h_{k,A}(Z_k)  \bigg\} \bigg]   \\
  =&  
  \argmin_{ g_{j}   \in \mathcal{H}_j^{(\beta_j)}, \beta_j \in \Theta,  h_k \in \mathcal{H}_k  }  
    \mathbb{E}\left[  
    \bigg\{ Y -  \sum_{j=1}^p g_{j,A}(\langle X_j, \beta_j \rangle) - \sum_{k=1}^q h_{k,A}(Z_k)  \bigg\}^2  
     +  2 \mu(\bm{X},\bm{Z}) \mathbb{E}\bigg[\sum_{j=1}^p g_{j,A}(\langle X_j, \beta_j \rangle) + \sum_{k=1}^q h_{k,A}(Z_k) | \bm{X},\bm{Z} \bigg] \right] \\ 
   =& 
    \argmin_{g_{j}   \in \mathcal{H}_j^{(\beta_j)}, \beta_j \in \Theta,  h_k \in \mathcal{H}_k }  
    \mathbb{E}\bigg[    \bigg\{ Y -  \sum_{j=1}^p g_{j,A}(\langle X_j, \beta_j \rangle) - \sum_{k=1}^q h_{k,A}(Z_k)  \bigg\}^2    \bigg],  
\end{aligned}
\end{equation*}
in which the second line follows from an application of the iterated expectation rule to condition on $(\bm{X}, \bm{Z})$ on the second term on the first line, and 
the third line follows from the constraint  imposed in (\ref{the.criterion}), that is, 
$\mathbb{E}[g_{j,A}(\langle X_j, \beta_j \rangle) | X_j ]  = 0, \ \forall \beta_j \in \Theta \ (j=1,\ldots,p)$ 
and $\mathbb{E}[h_{k,A}( Z_k) | Z_k ]  = 0 $ \ $(k=1,\ldots,q)$, 
which makes the second term on the second line of the above expression vanish to zero. 

Since the minimization in (\ref{the.criterion})  is in terms of  
$\{ g_{j}, \beta_j, h_k \}$,  
the right-hand side  
of (\ref{the.criterion}) 
can then be reduced to: 
 \begin{equation} \label{LS4}
\begin{aligned}
\{ g_{j}^\ast, \beta_{j}^\ast, h_{k}^\ast  \} 
\quad = \quad & \underset{ g_{j}   \in \mathcal{H}_j^{(\beta_j)}, \beta_j \in \Theta,  h_k \in \mathcal{H}_k }{\text{argmin}}
& &E \bigg[
 \bigg\{ Y -    \sum_{j=1}^p g_{j}(\langle X_j, \beta_j \rangle, A)  
- \sum_{k=1}^q h_{k}(Z_k,A) 
   \bigg\}^2 \bigg] 
    \\
& \text{subject to} & & E\left[g_{j}(\langle X_j, \beta_j \rangle,A) | X_j \right]  = 0 \quad 
\forall \beta_j \in \Theta \quad 
(j=1,\ldots,p) \quad \mbox{and}  \\
& 			    & & E\left[h_{k}( Z_k, A) | Z_k \right]  = 0 \quad (k=1,\ldots,q), 
\end{aligned}
\end{equation}
which is as appeared in the representation (3) of the main manuscript.

\subsection*{A.2. Proof of Theorem 1} \label{sec.proof.proposition1}
In this subsection, we provide the proof of Theorem 1 in Section 3.1 of the main manuscript. 
In order to simplify the exposition, 
we focus on the derivation of 
the minimizing functions 
$g_j \in \mathcal{H}_j^{(\beta_j)}$  $(j=1,\ldots,p)$ 
associated with the functional covariates $X_j$ $(j=1,\ldots, p)$, only. 
The minimizing functions $h_k \in \mathcal{H}_k$ $(k=1,\ldots, q)$ 
associated with the scalar covariates $Z_k$ $(k=1,\ldots, q)$ 
are derived in the similar way. 
Further, for fixed $\beta_j \in \Theta$ $(j=1,\ldots,p)$,   we write  $X_{\beta_j} = \langle X_j, \beta_j \rangle \in \mathbb{R}$ $(j=1,\ldots, p)$, for  notational simplicity.

The squared error criterion on the right-hand side of (\ref{LS4}) is 
\begin{equation} \label{LS.framework}
\begin{aligned}
 \mathbb{E} \bigg[ \big\{ Y -    \sum_{j=1}^p g_{j} ( X_{\beta_j},A)  
 \big\}^2  \bigg]  
 \propto  & \ \mathbb{E} \bigg[ Y  \sum_{j=1}^p g_{j}(X_{\beta_j},A)  
 -\big\{ \sum_{j=1}^p g_{j}(X_{\beta_j},A) 
  \big\}^2 /2
  \bigg] \quad  \mbox{(with respect to } \{g_{j}\} ) 
  \\
 =  & \ \mathbb{E} \bigg[ \big\{ \mu(\bm{X}) +   \sum_{j=1}^p g_{j}^\ast(X_{\beta_j^\ast},A) 
 \big\}
 \sum_{j=1}^p g_{j}(X_{\beta_j},A)  
- \big\{ \sum_{j=1}^p g_{j}(X_{\beta_j},A)  
 \big\}^2 /2
  \bigg]  \\
 = & \ \mathbb{E} \bigg[ \mu(\bm{X})  \sum_{j=1}^p g_{j}(X_{\beta_j},A) \bigg] +   \mathbb{E}\bigg[  \big\{ \sum_{j=1}^p g_{j}^\ast(X_{\beta_j^\ast},A)  
  \big\} \big\{ \sum_{j=1}^p g_{j}(X_{\beta_j},A)  
   \big\}   
  -\big\{ \sum_{j=1}^p g_{j}(X_{\beta_j},A)  
  \big\}^2 /2   \bigg]  \\
 = & \  \mathbb{E}\bigg[  \big\{ \sum_{j=1}^p g_{j}^\ast(X_{\beta_j^\ast},A)  
  \big\} \big\{ \sum_{j=1}^p g_{j}(X_{\beta_j},A)  
   \big\}   
  -\big\{ \sum_{j=1}^p g_{j}(X_{\beta_j},A)  
  \big\}^2 /2   \bigg], 
\end{aligned} 
\end{equation}
where the last equality follows from the constraint $ \mathbb{E}[g_{j}( X_{\beta_j},A) | X_j ]  = 0$ $(j=1,\ldots,p)$ in (\ref{LS4}) imposed on $\{g_j\}$, 
which implies  
$\mathbb{E} \big[  \mu( \bm{X} ) \big\{ \sum_{j=1}^p g_{j}\big(X_{\beta_j},A\big) \big\} \big]
=
\mathbb{E} \big[  \mathbb{E} \big[ \mu( \bm{X} ) \big\{\sum_{j=1}^p g_{j}(X_{\beta_j},A) \big\} \mid \bm{X} \big]  \big]
= 
\mathbb{E} \big[ \mu(\bm{X} ) \sum_{j=1}^p   \mathbb{E} \big[  g_{j}(X_{\beta_j},A)   \mid X_j \big]\big] =0$.  
From (\ref{LS.framework}), 
for fixed $\{\beta_j,  j=1,\ldots,p\}$, 
we can rewrite the squared error criterion in (\ref{LS4}) by (omitting the components associated with the scalar covariates):  
\begin{equation} \label{LS.framework2}
\underset{ \{ g_{j} \in \mathcal{H}_j^{(\beta_j)} \} }{\text{argmin}} \ \mathbb{E}\bigg[ \big(Y -    \sum_{j=1}^p g_{j}\big(X_{\beta_j},A\big)  \big)^2  \bigg] 
 \ =  \ 
  \underset{ \{ g_{j} \in \mathcal{H}_j^{(\beta_j)} \} }{\text{argmin}} \ \mathbb{E} \bigg[ \big( \sum_{j=1}^p g_{j}^\ast\big(X_{\beta_j^\ast},A\big)   -    \sum_{j=1}^p g_{j}\big(X_{\beta_j},A\big)  \big)^2  \bigg].  
  \end{equation} 
In the following, we closely follow the proof of Theorem 1 in \cite{SAM}. 
The Lagrangian in (4) of the main manuscript, for fixed $\{\beta_j,  j=1,\ldots,p\}$ 
can be rewritten as: 
\begin{equation}\label{H}
Q( \{g_j\}; \lambda)  := 
 \mathbb{E} \bigg[ \big( \sum_{j=1}^p g_{j}^\ast(X_{\beta_j^\ast},A)   -    \sum_{j=1}^p g_{j}(X_{\beta_j},A)  \big)^2  \bigg]
+ \lambda  \sum_{j=1}^p \lVert g_{j} \rVert.  
\end{equation}
Fixing $\{\beta_j,  j=1,\ldots,p\}$, 
 for each $j$, 
let us consider the minimization of (\ref{H}) 
with respect to the component function $g_{j} \in \mathcal{H}_j^{(\beta_j)}$, 
holding the other component functions 
$\{ g_{j'}, j' \ne j\}$ fixed. 
The stationary condition is obtained by setting its Fr\'{e}chet derivative to 0. 
Denote by 
 $\partial_{j} Q( \{g_j\}; \lambda; \eta_{j})$ 
 the directional derivative with respect to $g_{j} \in \mathcal{H}_j^{(\beta_j)}$ $(j=1,\ldots,p)$ 
 in an arbitrary direction which we denote by $\eta_{j} \in \mathcal{H}_j^{(\beta_j)}$. 
 Then, for fixed $\{\beta_j,  j=1,\ldots,p\}$, the stationary point of the Lagrangian (\ref{H}) can be formulated as:
\begin{equation}\label{stationary.cond}
 \partial_{j}  Q( \{g_j\}; \lambda; \eta_{j}) 
 = 2 \mathbb{E}\left[  (g_{j} - \widetilde{R}_{j} + \lambda \nu_{j}  ) \eta_{j}  \right] =0,  
\end{equation}
 where 
\begin{equation} \label{R.j}
\widetilde{R}_{j}  := \sum_{j=1}^p g_{j}^\ast(X_{\beta_j^\ast},A) - \sum_{j' \ne j} g_{j^\prime,A}(X_{\beta_{j^\prime}}), 
 \end{equation} 
representing the partial residual for the $j$th component function $g_{j}$, and the function 
 $\nu_{j}$ 
 is an element of the subgradient  
 $\partial  \lVert g_{j} \rVert $, 
 which satisfies 
 $\nu_{j} =   g_{j} /  \lVert g_{j} \rVert $ if $\lVert g_{j} \rVert  \ne 0$, 
 and 
  $\nu_{j} \in \{ s \in \mathcal{H}_j^{(\beta_j)}  \mid   \lVert s \rVert  \le 1 \} $, otherwise. 
Applying the iterated expectations to condition on $(X_{\beta_j}, A)$,  the stationary condition (\ref{stationary.cond}) can be rewritten as:  
\begin{equation} \label{stationary.cond2}
  2 \mathbb{E}\left[ \left(g_{j} -  \mathbb{E}\left[ \widetilde{R}_{j} | X_{\beta_j}, A \right]  + \lambda \nu_{j}   \right) \eta_{j}   \right] =0.  
\end{equation} 
Since the function 
 $g_{j}   -  \mathbb{E}\left[ \widetilde{R}_{j}  | X_{\beta_j}, A \right]  + \lambda \nu_{j}   \in \mathcal{H}_j^{(\beta_j)}$, 
 we can evaluate (\ref{stationary.cond}) (i.e., expression (\ref{stationary.cond2})) in the particular direction: 
$\eta_{j}  =  g_{j}   -  \mathbb{E}\left[ \widetilde{R}_{j}  | X_{\beta_j}, A \right]  + \lambda \nu_{j} $, 
which gives 
$ \mathbb{E}\left[ \left( g_{j}  -  \mathbb{E}\left[ \widetilde{R}_{j}  | X_{\beta_j}, A  \right]  + \lambda \nu_{j}  \right)^2 \right] =0.$ 
This equation implies: 
\begin{equation} \label{stationary.condition}
g_{j}  + \lambda \nu_{j} =  \mathbb{E}\left[ \widetilde{R}_{j} | X_{\beta_j}, A \right]    \quad 
 \mbox{(almost surely).}
\end{equation} 
Let $f_{j}$ denote  
the right-hand side of (\ref{stationary.condition}), i.e., 
$f_{j} (= f_{j}(X_{\beta_j},A)) := \mathbb{E}\left[ \widetilde{R}_{j}  | X_{\beta_j}, A \right]$. 
If $\lVert g_{j} \rVert  \ne 0$, 
then $\nu_{j}  = g_{j}/  \lVert g_{j} \rVert$.  
Therefore, by (\ref{stationary.condition}), we have 
$\lVert f_{j}  \rVert = \lVert  g_{j}+ \lambda  g_{j} /  \lVert g_{j} \rVert    \rVert 
= \lVert  g_{j}  \rVert + \lambda 
\ge  \lambda$.    
On the other hand, 
if $\lVert g_{j} \rVert  = 0$,  
then $g_{j} = 0$ (almost surely), 
and $\lVert \nu_{j} \rVert  \le 1$. 
Then, condition (\ref{stationary.condition}) implies that 
$\lVert f_{j} \rVert  \le \lambda$. 
This gives us the equivalence between 
$\lVert f_{j} \rVert  \le \lambda$ and the statement 
$g_{j}= 0$ (almost surely). 
Therefore, 
condition (\ref{stationary.condition}) leads to the following expression: 
\begin{equation*} \label{g.jt.solution}
\left( 1 + \lambda /  \lVert  g_{j}  \rVert  \right)  g_{j}   =  f_{j}  \quad 
\mbox{(almost surely)}
\end{equation*} 
if $\lVert f_{j} \rVert  > \lambda$, 
and $ g_{j}  = 0$ (almost surely), otherwise; this implies the soft thresholding update rule for $g_{j}$ appeared in (5) of the main manuscript. 

Now we will derive the expression (6) of the main manuscript for the function $f_j$. 
Note, the underlying model (\ref{the.model}) (if we omit the components associated with the scalar covariates) implies that 
 $ \sum_{j=1}^p g_{j}^\ast(X_{\beta_j^\ast},A)  = \mathbb{E}[Y | \bm{X}, A] - \mu(\bm{X})$. Thus, 
 (\ref{R.j}) can be equivalently written as: 
 $\widetilde{R}_{j} =  \mathbb{E}[Y | \bm{X}, A] - \mu(\bm{X})  - \sum_{j' \ne j} g_{j'}(X_{\beta_{j^\prime}},A).$
Therefore, the function $f_{j}(X_{\beta_j},A) = \mathbb{E}\left[ \widetilde{R}_{j}  | X_{\beta_j}, A \right]$ can be written as: 
\begin{equation*} \label{final.eq}
\begin{aligned}
 f_{j}(X_{\beta_j},A)  \ &=   \mathbb{E}\big[ \mathbb{E}[Y | \bm{X}, A] - \mu(\bm{X})  - \sum_{j' \ne j} g_{j'}(X_{\beta_{j^\prime}},A) \mid  X_{\beta_j}, A \big]  \\
 &= \mathbb{E}\big[ \mathbb{E}[Y | \bm{X}, A]  - \sum_{j' \ne j} g_{j'}(X_{j^\prime},A) \mid  X_{\beta_j}, A \big]   - \mathbb{E}\big[ \mu(\bm{X}) \mid  X_{\beta_j}, A \big] \\
 &= \mathbb{E}\big[ Y - \sum_{j' \ne j} g_{j'}(X_{\beta_{j^\prime}},A) \mid  X_{\beta_j}, A \big]   - \mathbb{E}\big[\mu(\bm{X}) \mid X_{\beta_j} \big]  \\
 &=  \mathbb{E}\big[ Y - \sum_{j' \ne j} g_{j'}(X_{\beta_{j^\prime}},A) \mid  X_{\beta_j}, A \big]    - \mathbb{E}\big[\mu(\bm{X}) + \sum_{j=1}^p g_{j}^\ast\big(X_{\beta_j^\ast},A\big) \mid  X_{\beta_j} \big]   \\
     &= \mathbb{E}\big[ Y - \sum_{j' \ne j} g_{j'}(X_{\beta_{j^\prime}},A) \mid  X_{\beta_j}, A \big]   - \mathbb{E}\big[Y  \mid  X_{\beta_j}\big] \\
     &= \mathbb{E}\big[ Y - \sum_{j' \ne j} g_{j'}(X_{\beta_{j^\prime}},A) \mid  X_{\beta_j}, A \big]   - \mathbb{E}\big[Y  - \sum_{j' \ne j} g_{j'}(X_{\beta_{j^\prime}},A) \mid  X_{\beta_j} \big] \\ 
     &= \mathbb{E}\big[ R_{j} \mid  X_{\beta_j}, A \big]   - \mathbb{E}\big[R_{j}  \mid X_{\beta_j} \big],  
\end{aligned} 
 \end{equation*}
 where the fourth equality of the expression follows from the identifiability constraint (\ref{the.condition}) of the underlying model (\ref{the.model}), 
 and the sixth equality follows from the  optimization constraint 
 $ \mathbb{E}[g_{j^\prime}( X_{\beta_{j^\prime}},A) | X_j ]  = 0$ $(j^\prime \ne j)$ implied by (\ref{LS4}) imposed on $\{g_{j^\prime}, j^\prime \ne j  \}$;   
this gives the  expression  (6) of the main manuscript  for $f_j$.

\subsection*{A.3. Description of general linear smoothers for the component functions} \label{sec.fff}

As a remark to Section 3.2.1 of the main manuscript, we note that 
any  scatterplot smoother 
 can be utilized to obtain 
 the sample counterpart (16) of the main manuscript of the coordinate-wise solution  (5) for the component functions $g_j$, i.e., estimation of the component functions $g_j$ is not restricted to regression splines.  

To estimate the function $f_j$ in (6), 
we can estimate the system of treatment $a$-specific functions 
$\mathbb{E}[R_{j} | \langle \widehat{\beta}_j, X_{ij} \rangle, A=a ]$  $(a=1,\ldots,L)$ 
(which corresponds to the first term on the right-hand side of  (6)  if we fix $\beta_j = \widehat{\beta}_j$), 
by performing separate nonparametric regressions of $\widehat{R}_j$ on regressor $\langle \widehat{\beta}_j, X_{ij} \rangle$ separately for each treatment condition $A=a$ $(a=1,\ldots,L)$. 
We can also estimate the function $- \mathbb{E}[R_{j} | \langle \beta_j, X_{ij} \rangle]$ (which 
corresponds to the second term $- \mathbb{E}[R_{j} | \langle \beta_j, X_{ij} \rangle]$ on the right-hand side of  (6) if we fix $\beta_j = \widehat{\beta}_j$), 
by performing a nonparametric regression of $\widehat{R}_j$ on regressor $\langle \widehat{\beta}_j, X_{ij} \rangle$. 
 Adding these two function estimates provides an estimate for $f_j$ in  (6). 
  Evaluating this estimate of $f_j$ at the $n$ points $( \langle \widehat{\beta}_j, X_{ij} \rangle, A_i)$ $(i=1,\ldots,n)$
 gives an estimate $\widehat{\bm{f}}_j \in \mathbb{R}^n$ in (16). Then we can compute the corresponding soft-threshold  estimate $\widehat{\bm{g}}_j\in \mathbb{R}^n$ and conduct the coordinate descent procedure described in Algorithm~1 of the main manuscript.

\subsection*{A.4. Supplementary information for Section 3.2.1} \label{sec.eee}

The restriction of the function  $g_j$ to the form 
(12) of the main manuscript 
 restricts also the minimizing function $g_j$  in (5) of the main manuscript \big(note, $g_{j}( \langle X_{j}, \widehat{\beta}_{j} \rangle, A) = s_j^{(\lambda)} f_{j}( \langle X_{j}, \widehat{\beta}_{j} \rangle, A)$, where $s_j^{(\lambda)} = \left[ 1- \lambda/ \lVert f_{j} \rVert  \right]_{+}$\big) to the form (12) of the main manuscript.  
In particular, 
we can express the function $f_j$ in (6) 
of the main manuscript as: 
 \begin{equation} \label{eq.4}
\begin{aligned}
 f_{j,A}(\langle X_{j}, \widehat{\beta}_{j} \rangle)  \
&=   \  \mathbb{E}[R_j | \langle X_{j}, \widehat{\beta}_{j} \rangle, A ] -   \sum_{a=1}^L \pi_a  \mathbb{E}[R_{j} |  \langle X_{j}, \widehat{\beta}_{j} \rangle, A=a ]  \\
&=  \ \bm{\Psi}_j(\langle X_{j}, \widehat{\beta}_{j} \rangle)   \bm{\theta}_{j,A}^\ast  -   \bm{\Psi}_j(\langle X_{j}, \widehat{\beta}_{j} \rangle)  \{ \sum_{a=1}^L \pi_a  \bm{\theta}_{j,a}^\ast \}, 
\end{aligned}
\end{equation} 
where $\{ \bm{\theta}_{j,a}^\ast \}_{a \in \{1,\ldots,L\}}   := \underset{ \{ \bm{\theta}_{j,a} \in \mathbb{R}^{d_j}\}_{a \in \{1,\ldots,L\}} }{\text{argmin}}  
  \mathbb{E}\left[ \big\{ R_j -  \bm{\Psi}_j( \langle X_{j}, \widehat{\beta}_{j} \rangle)^\top   \bm{\theta}_{j,A} \big\}^2\right]$.  
  In (\ref{eq.4}), 
the first term,  
$\bm{\Psi}_j(\langle X_{j}, \widehat{\beta}_{j} \rangle)   \bm{\theta}_{j,A}^\ast$, 
corresponds  to the  $L^2$ projection of the $j$th partial residual $R_j$ in (7) (of the main manuscript) 
onto the
class of functions of the form 
  (12) (without the imposition of the constraint (13), that is, the constraint $\sum_{a=1}^{L} \pi_a \bm{\theta}_{j,a} = \bm{0}$), 
whereas the second term,  
$-   \bm{\Psi}_j(\langle X_{j}, \widehat{\beta}_{j} \rangle)  \{ \sum_{a=1}^L \pi_a  \bm{\theta}_{j,a}^\ast \}$, 
  simply centers the first term to satisfy the linear constraint,  
 $\sum_{a=1}^{L} \pi_a \bm{\theta}_{j,a} = \bm{0}$. 
 Then it follows that $f_j$, as given in  (\ref{eq.4}), corresponds to the $L^2$ projection of $R_j$ 
 onto the subspace of  measurable functions of the form (12) 
 subject to the linear constraint (13) of the main manuscript.

\subsection*{A.5. Simulation results under a ``linear'' $A$-by-$(\bm{X},\bm{Z})$  interaction effect scenario} \label{sec.ggg}

In this subsection, 
as an extension of the simulation example in Section 4.1 of the main manuscript,  
we consider a case where the treatment effect varies \textit{linearly} in the covariates $(\bm{X},\bm{Z})$, i.e., a ``\textit{linear}'' $A$-by-$(\bm{X},\bm{Z})$ interaction effect scenario 
and assess the  ITR estimation 
performance of the methods. 
Specifically, we consider the data generation model: 
\begin{equation} \label{sim.model1}
\begin{aligned}
Y_i &=   
\delta\bigg\{ \sum_{j=1}^8 \sin( \langle \eta_j, X_{ij} \rangle ) + \sum_{k=1}^8 \sin(Z_{ik}) \bigg\}
 + \\
&
4(A_i - 1.5) 
\bigg[  \langle \beta_1, X_{i1} \rangle/1.5   -  \langle \beta_2, X_{i2} \rangle/1.5  
+ Z_{i1} /1.5- Z_{i2} /1.5
+ \xi  \big\{ \langle X_{i1}, X_{i2} \rangle/1.5 
  + Z_{i1} Z_{i2}/1.5 \big\} 
 \bigg]   +\epsilon_i, 
\end{aligned}
\end{equation}
in which, when $\xi = 0$, 
a functional \textit{linear} model specifies 
the $A$-by-$(\bm{X},\bm{Z})$ interaction effect term (i.e., the second term on the right-hand side of (\ref{sim.model1})). 
However, when $\xi=1$, 
the underlying model (\ref{sim.model1}) deviates from the exact linear  $A$-by-$(\bm{X},\bm{Z})$ interaction effect structure, 
and  in such a case, the model CFAM-lin (as well as CFAM) is misspecified. 
The contribution 
to the variance of $Y$ from the main and the interaction effect terms   
in (\ref{sim.model1}) 
was made similar to that of 
the data generating 
model (25) of Section 4.1 of the main manuscript.

\begin{figure} [H]
\begin{center}
\begin{tabular}{c}
\includegraphics[width=6.4in, height = 1.8 in]{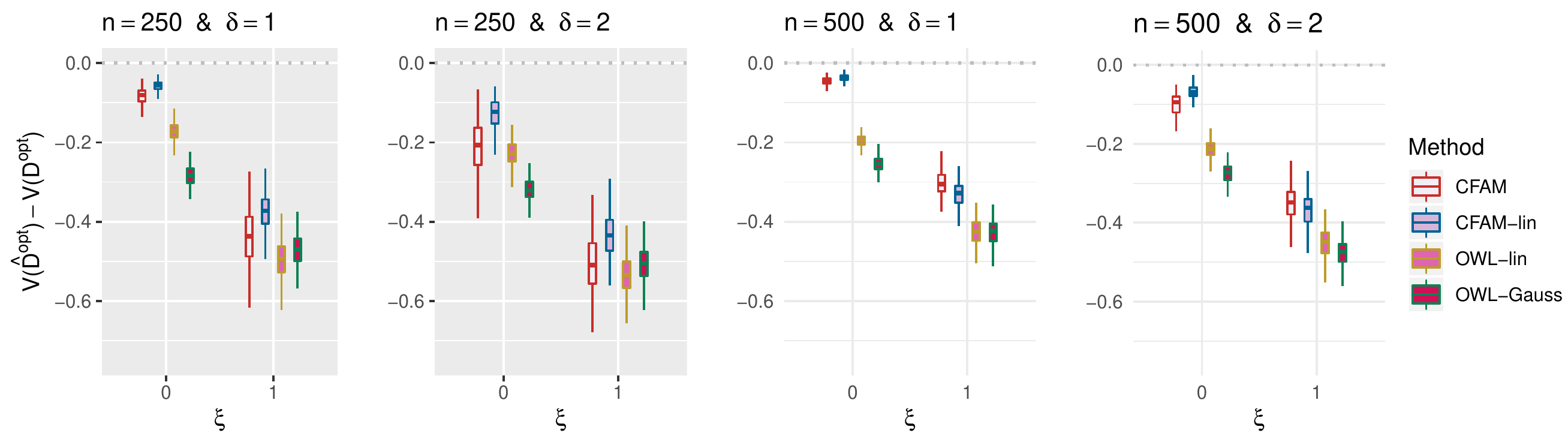}
\end{tabular}
\end{center}
\caption{
Boxplots obtained from  $200$ Monte Carlo simulations 
comparing $4$ approaches to estimating $\mathcal{D}^{opt}$, 
given each scenario indexed by $\xi \in \{0,1\}$, $\delta \in \{1,2\}$ and  $n\in \{250, 500\}$. 
The dotted horizontal line represents the optimal value corresponding to $\mathcal{D}^{opt}$.   
} \label{fig.sim.result2}
\end{figure}

 Figure \ref{fig.sim.result2} illustrates the boxplots, obtained from $200$ simulation runs, 
 of the normalized values $V(\widehat{\mathcal{D}}^{opt})  - V(\mathcal{D}^{opt})$ (normalized by the optimal values $V(\mathcal{D}^{opt})$) 
  of the decision rules  $\widehat{\mathcal{D}}^{opt}$ 
estimated from the four approaches described in Section 4.1 of the main manuscript, 
for each combination of $n \in \{250, 500 \}$, 
$\xi \in \{0, 1\}$ (corresponding to   \textit{correctly-specified} or \textit{mis-specified} CFAM scenarios, respectively) and 
 $\delta \in \{1, 2 \}$  (corresponding to \textit{moderate} or  \textit{large} main effects, respectively). 
 
 In all scenarios with $\xi = 0$  (i.e., when the linear interaction model is correctly specified), 
CFAM-lin outperforms CFAM, 
but by a relatively small amount in comparison to the difference in performance appearing  
in Figure 1 of the main manuscript, in which CFAM outperforms CFAM-lin. 
Moreover,  if the underlying model deviates from the exact linear structure (i.e., $\xi = 1$ in model (\ref{sim.model1})) 
 and $n=500$, 
 the more flexible CFAM tends to outperform CFAM-lin.  
Given the outstanding performance of  CFAM compared to CFAM-lin  in the \textit{nonlinear} $A$-by-$(\bm{X},\bm{Z})$  interaction effect scenarios  considered in the main manuscript, this result suggests that, in the absence of prior knowledge about the form of the interaction effect, flexible modeling of the interaction effect using the proposed CFAM  can lead to good results in comparison to  CFAM-lin.

\subsection*{A.6. Separate modeling of the $(\bm{X}, \bm{Z})$ ``main'' effect component} \label{sec.A.6.}

Under model (\ref{the.model}), constraint (\ref{the.condition}) (i..e, constraint (2) of the main manuscript) ensures that 
$$E\left[ 
\mu(\bm{X}, \bm{Z}) 
\left\{ 
\sum_{j=1}^p g_{j}(\langle  X_j, \beta_{j}  \rangle, A ) 
+ 
\sum_{k=1}^q h_{k}(Z_k, A) 
\right\}
\right] = 
E\left[ 
\mu(\bm{X}, \bm{Z}) 
E\left\{ 
\sum_{j=1}^p g_{j}(\langle X_j, \beta_{j} \rangle, A ) 
+ 
\sum_{k=1}^q h_{k}(Z_k, A) 
\mid  \bm{X}, \bm{Z} 
\right\}
\right]
= 0,
$$ 
where, on the right-hand side, we apply the iterated expectation rule to condition on $(\bm{X}, \bm{Z})$, 
which implies:  
\begin{equation}\label{orthogonality}
\mu(\bm{X}, \bm{Z}) 
\quad 
\perp
\quad 
\sum_{j=1}^p g_{j}(\langle  X_j, \beta_{j} \rangle, A ) 
+ 
\sum_{k=1}^q h_{k}(Z_k, A)
\end{equation}
in $L^2$. 
The orthogonality (\ref{orthogonality}) 
conceptually and also practically 
implies that, under the squared error minimization criterion,
 the optimization for $\mu$ and the components $\{ g_j, \beta_j, h_k  \}$ in model (\ref{the.model}) (subject to (\ref{the.condition}))  can be performed separately, without iterating between the two optimization procedures. 
To be specific, we can solve for the $(\bm{X},\bm{Z})$ ``main'' effect:
\begin{equation}\label{main}
\mu^\ast 
\quad = \quad \underset{ \mu  \in \mathcal{H} }{\text{argmin}} \quad 
E\left[ \left\{ Y - \mu(\bm{X}, \bm{Z}) \right\}^2 \right],
\end{equation}
and can separately solve 
for the $(\bm{X},\bm{Z})$-by-$A$ interaction effect via optimization 
(\ref{LS4}). 
In optimization (\ref{main}), $\mathcal{H}$ represents a (possibly misspecified) $L^2$ space of functionals over $(\bm{X}, \bm{Z})$. 
Even if  the true $\mu$ in (\ref{the.model}) is not in the  class $\mathcal{H}$, the representation (\ref{LS4}) that specifies the optimal $(\bm{X},\bm{Z})$-by-$A$  interaction effect components, i.e., $\{ g_{j}^\ast, \beta_{j}^\ast, h_{k}^\ast  \}$,   
is not affected by the possible misspecification for $\mu$, due to the orthogonality (\ref{orthogonality}).

For the case of a continuous outcome $Y$, 
\cite{MC} (in the  linear regression context with scalar covariates and Lasso regularization) 
and \cite{CSIM} (in the single-index regression  context  with scalar covariates) 
proposed to separately model and
 fit the main effect component $\mu^\ast$, 
by leveraging the orthogonality property analogous to (\ref{orthogonality}),  
 and then utilize the residualized outcome $Y - \widehat{\mu}^\ast(\bm{X},\bm{Z})$ (instead of using the original outcome $Y$) 
for the estimation of the interaction effect components.  
This residualization procedure that utilizes separately fitted main effect  was termed \textit{efficiency augmentation} by \cite{MC},  and can improve the efficiency of the estimator for the interaction effect components (while maintaining the consistency of the estimator). 
In what follows, we illustrate an additional set of simulations supplementing the results of Section 4.1 of the main manuscript, 
which  demonstrates  some performance improvement of the CFAM method via an efficiency augmentation procedure. 

Under the simulation model (25) of the main manuscript (in which the corresponding results are reported in Table 1 of the main manuscript) for generating the data, 
 we report additional simulation results associated with the CFAM method when the $(\bm{X}, \bm{Z})$  ``main'' effect component of the data generating model (25)  
 is modeled by a functional additive regression, i.e., by the model:  
 $\mu(\bm{X}, \bm{Z}) = \sum_{j=1}^p \widetilde{g}_j(\langle  X_j, \widetilde{\beta}_j \rangle) 
 + \sum_{k=1}^q \widetilde{h}_k(Z_k)$, 
 estimated based on an $L^1$ regularization that is similar to (4) of the main manuscript, 
   with the associated
  tuning parameters selected as in the CFAM method) and 
the corresponding residualized outcome is used to  
implement the CFAM method.

 In Table~\ref{t:two}, those rows with  the label  
``CFAM'' 
correspond to what are reported  
in Table 1 of the main manuscript, 
whereas those with 
``CFAM($\mu$)'' correspond to  the cases where  that  ``main'' effect component $\mu$ is modeled by the functional additive regression model described above. 
  As in Table 1 of the main manuscript, we report the root squared error 
   $\mbox{RSE}(\beta_j) = \sqrt{ \int (\widehat{\beta}_j(s) - \beta_j(s))^2 ds }$ $(j=1,2)$, 
  where the parameters  $\beta_1$ and $\beta_2$ are given in the data model (25) of the main manuscript, 
and 
 $\widehat{\beta}_1$ and $\widehat{\beta}_2$ are the corresponding estimates. 
In addition to RSE, we also report the optimal ITR estimation performance of CFAM and CFAM($\mu$), in terms of the (normalized) value $V^\ast(\widehat{\mathcal{D}}^{opt}) = V(\widehat{\mathcal{D}}^{opt}) - V(\mathcal{D}^{opt})$ 
  (where a larger value of $V^\ast(\widehat{\mathcal{D}}^{opt})$ is desired).

\begin{table} 
\caption{
Comparison of the performance of CFAM and CFAM($\mu$), with respect to 
the parameter estimation assessed by 
 the root squared error  $\mbox{RSE}(\beta_j)$ (a smaller value of $\mbox{RSE}(\beta_j)$ is desired) and the optimal ITR estimation assessed by  
$V^\ast(\widehat{\mathcal{D}}^{opt}) = V(\widehat{\mathcal{D}}^{opt}) - V(\mathcal{D}^{opt})$ (a larger value of $V^\ast(\widehat{\mathcal{D}}^{opt})$ is desired), 
for varying $\delta \in \{1, 2 \}$ and $n \in \{250, 500, 1000\}$.  
} \label{t:two}
\begin{center}
\begin{tabular}{llrrrrrr}
\hline
	  &	 & \multicolumn{3}{c}{$\delta =1$ (\textit{Moderate} ``main'' effect)}  &   \multicolumn{3}{c}{$\delta=2$ (\textit{Large} ``main'' effect)} \\ \hline
   &	   & \multicolumn{1}{c}{$n=250$} &  \multicolumn{1}{c}{$n=500$} & \multicolumn{1}{c}{$n=1000$} &   \multicolumn{1}{c}{$n=250$} &  \multicolumn{1}{c}{$n=500$} & \multicolumn{1}{c}{$n=1000$} \\ \hline
 & $\mbox{RSE}(\beta_1)$ & 0.53(0.08) & 0.34(0.02) & 0.26(0.02) & 0.60(0.14) & 0.38(0.05)  &  0.29(0.03)   \\
CFAM & $\mbox{RSE}(\beta_2)$ & 0.53(0.06) & 0.34(0.02) & 0.27(0.01)   &  0.59(0.13) & 0.39(0.07)  & 0.29(0.03)  \\
 & $V^\ast(\widehat{\mathcal{D}}^{opt}) $ & -0.07(0.04) & -0.03(0.01) & -0.01(0.01) & -0.16(0.07) & -0.07(0.02)  & -0.04(0.01)   \\
\hline
 & $\mbox{RSE}(\beta_1)$ & 0.52(0.06) & 0.33(0.02) & 0.26(0.01) & 0.57(0.13) & 0.36(0.03)  &   0.28(0.02)  \\
CFAM($\mu$) & $\mbox{RSE}(\beta_2)$ & 0.52(0.06) & 0.33(0.01) & 0.26(0.01)   &  0.56(0.11) & 0.36(0.04)  &  0.28(0.02) \\
 & $V^\ast(\widehat{\mathcal{D}}^{opt})$ &-0.06(0.04) & -0.02(0.01) &  -0.01(0.00)  & -0.12(0.05) & -0.05(0.01)  & -0.02(0.01)  \\
\hline
\end{tabular}
\end{center}
\end{table}

The results in Table~\ref{t:two} indicate that, especially when $\delta = 2$ (i.e, for the large ``main'' effect cases),   
the efficiency augmentation procedure based on the functional additive regression model for the $(\bm{X},\bm{Z})$ ``main'' effect 
improves  CFAM (comparing the rows that are labeled as CFAM vs. those of CFAM($\mu$)), 
  in terms of both the parameter estimation performance, i.e., $\mbox{RSE}(\beta_j)$ $(j=1,2)$   
and  the optimal ITR estimation performance, i.e., $V^\ast(\widehat{\mathcal{D}}^{opt})$.


\end{document}